\documentclass[aps,prd,twocolumn,preprintnumbers,nofootinbib,amsmath,amssymb,superscriptaddress,10pt,floatfix]{revtex4-2}

\usepackage{graphicx}
\usepackage{placeins}
\graphicspath{{figs/}}

\usepackage{hyperref}
\usepackage[capitalise]{cleveref}
\crefname{section}{Sec.}{Sections} 

\usepackage[font=small, labelfont=bf, textfont={small,it}]{caption}
\usepackage{multirow}
\usepackage{makecell}
\usepackage{booktabs}
\usepackage{subcaption}

\usepackage{xspace}
\usepackage{cancel}
\usepackage[normalem]{ulem}
\usepackage{braket}
\usepackage[table]{xcolor}
\usepackage{orcidlink}

\newcommand{\be}{\begin{equation}}
\newcommand{\ee}{\end{equation}}

\newcommand{\SU}[1]{$\mathrm{SU}(#1)$\xspace}
\newcommand{\SUN}{$\mathrm{SU}(N)$\xspace}

\newcommand{\s}{\mathrm{s}}

\newcommand{\TBC}{TBCs\xspace}
\newcommand{\PBC}{PBCs\xspace}

\newcommand{\ov}{\scriptscriptstyle{\mathrm{OR}}}
\newcommand{\clov}{\scriptscriptstyle{\mathrm{clov}}}
\newcommand{\meas}{\scriptscriptstyle{\mathrm{meas}}}
\newcommand{\Tr}{\mathrm{Tr}}
\newcommand{\TGF}{\scriptscriptstyle{\mathrm{TGF}}}
\newcommand{\pt}{{\scriptscriptstyle{\rm pt}}}

\newcommand{\muhad}{\mu_{\scriptscriptstyle{\rm had}}}
\newcommand{\mupt}{\mu_{\scriptscriptstyle{\rm pt}}}

\newcommand{\LambdaMS}{\Lambda_{\scriptscriptstyle{\rm \overline{MS}}}}
\newcommand{\LambdaTGF}{\Lambda_{\scriptscriptstyle{\rm TGF}}}
\newcommand{\lambdaMS}{\lambda_{\scriptscriptstyle{\rm \overline{MS}}}}
\newcommand{\lambdaTGF}{\lambda_{\scriptscriptstyle{\rm TGF}}}
\newcommand{\lambdazeroTGF}{\lambda^{\scriptscriptstyle{(0)}}_{\scriptscriptstyle{\rm TGF}}}
\newcommand{\dd}{\mathrm{d}}
\newcommand{\eee}{\mathrm{e}}
\newcommand{\ii}{\mathrm{i}}
\newcommand{\MStext}{\overline{\mathrm{MS}}}

\begin{document}

\title{The large-\texorpdfstring{$N$}{N} Yang--Mills \texorpdfstring{$\Lambda$}{Lambda}-parameter from step scaling}

\author{Claudio Bonanno}
\email{claudio.bonanno@unibe.ch}
\affiliation{Albert Einstein Center for Fundamental Physics, Institute for Theoretical Physics, University of Bern, Sidlerstra{\ss}e 5, CH-3012 Bern, Switzerland}
\affiliation{Instituto de F\'isica Te\'orica UAM-CSIC, c/ Nicol\'as Cabrera 13-15, Universidad Aut\'onoma de Madrid, Cantoblanco, E-28049 Madrid, Spain}

\author{Jorge Luis Dasilva Gol\'an}
\email{jgolandas@bnl.gov}
\affiliation{Physics Department, Brookhaven National Laboratory, Upton, New York 11973, USA}

\author{Margarita Garc\'ia P\'erez}
\email{margarita.garcia@csic.es}
\affiliation{Instituto de F\'isica Te\'orica UAM-CSIC, c/ Nicol\'as Cabrera 13-15, Universidad Aut\'onoma de Madrid, Cantoblanco, E-28049 Madrid, Spain}

\author{Andrea Giorgieri}
\email{andrea.giorgieri@phd.unipi.it}
\affiliation{Dipartimento di Fisica, Università di Pisa \& INFN, Sezione di Pisa, Largo Pontecorvo 3, I-56127 Pisa, Italy}

\date{\today}

\begin{abstract}
We use the step-scaling method and results obtained at $N = 3, 5$ and $8$ to determine the $N$-dependence of the dynamically generated scale $\Lambda$ of $\mathrm{SU}(N)$ Yang--Mills theories. We implement the step-scaling method in a suitable finite-volume renormalization scheme based on twisted boundary conditions, introduced to effectively achieve large-$N$ volume independence, and on a coupling defined through the gradient flow. In the $\overline{\mathrm{MS}}$ scheme, we obtain the following values in terms of the gradient flow scale $t_0$: $\sqrt{8t_0}\Lambda_{\scriptscriptstyle{\overline{\mathrm{MS}}}} = 0.577(23)$, $0.632(32)$, and $0.611(43)$ for $N=3,5$ and $8$, respectively. They extrapolate to a large-$N$ value of: $\sqrt{8t_0}\Lambda_{\scriptscriptstyle{\overline{\mathrm{MS}}}} (N=\infty) = 0.639(36)$, and the $N$-dependence is given by $\sqrt{8t_0}\Lambda_{\scriptscriptstyle{\overline{\mathrm{MS}}}}(N)=0.639(36)[1-0.85(62)/N^2+\mathcal{O}(1/N^4)]$. This work represents the first calculation of the Yang--Mills $\Lambda$-parameter in the large-$N$ limit that does not rely on asymptotic scaling strategies.
\end{abstract}

\maketitle

\section{Introduction}\label{sec:intro}

The large-$N$ limit of four-dimensional \SUN non-Abelian gauge theories has long provided a valuable theoretical framework for understanding the non-perturbative dynamics of strong interactions~\cite{tHooft:1973alw,Witten:1978bc}. In this limit, gauge theories acquire a simplified diagrammatic structure, and a number of important features emerge naturally, including large-$N$ factorization, the master-field picture, and volume independence. Although $N=3$ is the physically relevant case, the possibility of performing a systematic $1/N$ expansion around $N=\infty$ provides an important tool for connecting the large-$N$ and the real-world theories. Thus, the large-$N$ limit not only allows one to probe confinement and asymptotic freedom from a broader perspective, but also provides a useful framework to study several interesting non-perturbative aspects of hadron phenomenology, with $\eta^\prime$ meson physics and the chiral anomaly being paramount examples~\cite{tHooft:1976rip,Witten:1979vv,Veneziano:1979ec,Kawarabayashi:1980dp,Witten:1980sp,DiVecchia:1980yfw}.

Due to its theoretical significance and phenomenological implications for non-perturbative gauge dynamics, the large-$N$ limit of Yang--Mills theories has been extensively studied from first principles by means of numerical Monte Carlo simulations on the lattice, see e.g., Refs.~\cite{Lucini:2012gg,Lucini:2013qja,Hernandez:2020tbc,GarciaPerez:2020gnf}. In this context, a particularly relevant quantity is the so-called $\Lambda$-parameter, which sets the dynamical scale of the theory through dimensional transmutation. Determining its dependence on the number of colors is interesting in its own right because it probes the $N$-dependence of the confinement scale and also provides a clean setting in which to compare different non-perturbative strategies for extracting renormalization-group invariants.

Unlike the case of $N=3$, all large-$N$ studies of $\Lambda$~\cite{Lucini:2001ej,Allton:2008ty,Lohmayer:2012ue,Gonzalez-Arroyo:2012euf,Athenodorou:2021qvs,Butti:2023hfp} have so far relied on asymptotic-scaling techniques, whereas in this work we employ a finite-volume \emph{step-scaling strategy}~\cite{Luscher:1991wu}, based on the gradient flow~\cite{Narayanan:2006rf,Lohmayer:2011si,Luscher:2009eq} and twisted boundary conditions~\cite{tHooft:1979rtg} (for an exploratory study of $\Lambda$ using step scaling in \SU{4}, see~\cite{Lucini:2008vi}). The conceptual differences between these two approaches are important. In asymptotic-scaling analyses, one infers the $\Lambda$-parameter from the dependence of an ultra-violet (i.e., short-distance) observable on the bare coupling, assuming that the running scale is given by $\mu = 1/a$ (with $a$ the lattice spacing), and that the running of $a$ with the bare coupling (either the standard or an improved one) can be described by the expected perturbative expansion. Asymptotic scaling requires both large volumes and very small lattice spacings, leading to a window problem, since achieving both conditions in a single simulation is prohibitively expensive. In contrast, the step-scaling method determines the running of a renormalized coupling non-perturbatively through a sequence of finite-volume calculations, recursively connecting a low-energy hadronic scale to high-energy scales, where perturbation theory becomes reliable. This removes the need to use large volumes, and instead exploits the smallness of the box in physical units as a means to run the coupling up to the perturbative regime, making step scaling particularly well suited for controlled determinations of $\Lambda$.

The $\Lambda$-parameter is a renormalization-group invariant, meaning that it does not depend on the energy scale at which it is defined, but it is scheme dependent and therefore its determination requires the choice of a suitable renormalization scheme. In this work we adopt the \emph{Twisted Gradient Flow} scheme~\cite{Ramos:2014kla,Bribian:2019ybc,Bribian:2021cmg}, which combines three ingredients: a coupling defined through the gradient flow, the use of twisted boundary conditions, and an asymmetric space-time volume $\ell^2\times\ell_s^2$, with $\ell = N\ell_s$. This framework, theoretically based on the concept of \emph{twisted volume reduction}~\cite{PhysRevLett.48.1063,Gonzalez-Arroyo:1982hwr,Gonzalez-Arroyo:1982hyq,Gonzalez-Arroyo:2010omx,Gonzalez-Arroyo:2014dua}, is particularly attractive for large-$N$ studies since, in perturbation theory, the effective size that controls finite-volume effects is $\ell \gg \ell_s$, with the resulting setup having a reduced memory footprint compared with more conventional symmetric volumes.

Due to its intrinsic theoretical interest, the $\Lambda$-parameter of the pure-gauge theory has been extensively studied in the literature in the past~\cite{Alles:1996ka,CAPITANI1999669,Boucaud:1998xi,Boucaud:1998bq,Becirevic:1999hj,Becirevic:1999uc,Boucaud:2000nd,Boucaud:2000ey,DeSoto:2001qx,Boucaud:2001st,PhysRevD.73.014513,Boucaud:2005gg,Boucaud:2008gn,Sternbeck:2009hna,Ilgenfritz:2010gu,Brambilla:2010pp,Sternbeck:2012qs,Gonzalez-Arroyo:2013bta,Kitazawa:2016dsl,Ishikawa:2017xam,Husung:2017qjz}. More recently, this quantity has acquired renewed phenomenological relevance because it is an essential input for decoupling-based strategies~\cite{DallaBrida:2019mqg} that employ non-perturbative pure-gauge results to accurately extract the strong coupling in full QCD~\cite{DallaBrida:2022eua,DallaBrida:2026kuo} (see also~\cite{DallaBrida:2020pag,DelDebbio:2021ryq,dEnterria:2022hzv} for recent reviews on the topic). Thus, an accurate determination of the \SU{3} pure-gauge $\Lambda$-parameter has been the target of several recent works~\cite{DallaBrida:2019wur,Nada:2020jay,Husung:2020pxg,Bribian:2021cmg,Athenodorou:2021qvs,Hasenfratz:2023bok,Wong:2023CY,PhysRevD.109.114517} (see also the reviews~\cite{FlavourLatticeAveragingGroupFLAG:2021npn,FlavourLatticeAveragingGroupFLAG:2024oxs}). From this point of view, understanding the large-$N$ behavior of $\Lambda$ and comparing different methods for its determination is not only of conceptual value, but also helps to clarify the systematics of modern precision studies, especially in light of the existing tensions between results obtained using step-scaling and asymptotic-scaling approaches, see the discussion in Refs.~\cite{FlavourLatticeAveragingGroupFLAG:2021npn,FlavourLatticeAveragingGroupFLAG:2024oxs}. The present work should therefore be seen not only as a study of large-$N$ Yang--Mills dynamics, but also as an alternative setting in which possible differences between these methods can be investigated.

The paper is organized as follows. In \cref{sec:strategy} we provide a general overview of the theoretical framework underlying our step-scaling calculation in the continuum theory, as well as a brief discussion about the $\Lambda$-parameter and perturbation theory. In \cref{sec:lattice} we describe our lattice setup, namely, the adopted lattice action and simulation algorithms, and the chosen discretization for all relevant observables. In \cref{sec:results} we present our lattice implementation of the step-scaling method and our numerical results. These include the determination of the step-scaling function, the extraction of the $\Lambda$-parameter in the $\MStext$ scheme in units of a reference hadronic scale, and its conversion to customary units in terms of the gradient-flow scale $t_0$. In \cref{sec:conclusions}, given that this is the first large-$N$ calculation of $\LambdaMS$ with step scaling, we also provide a comprehensive, critical comparison of our determinations for $\LambdaMS(N)\sqrt{8t_0}$ with previous results in the literature. Finally, we draw our conclusions and discuss future outlooks of this work.

\section{The \texorpdfstring{$\Lambda$}{Lambda}-parameter from step scaling in the twisted gradient flow scheme in the continuum theory}\label{sec:strategy}

In this section, we outline the strategy used to determine the $\Lambda$-parameter in the continuum, while its practical lattice implementation is described in the following sections. We begin by recalling the definition of $\Lambda$ and its relation to the renormalization group flow, and then describe how it can be computed non-perturbatively using the \emph{step-scaling} method~\cite{Luscher:1991wu}. Finally, we introduce the renormalization scheme adopted in this work, namely the \emph{Twisted Gradient Flow} (TGF) scheme~\cite{Ramos:2014kla,Bribian:2019ybc,Bribian:2021cmg}.

\subsection{The step-scaling method}\label{sec:step-scaling}

Consider the Gell-Mann--Low $\beta$-function defined in a renormalization scheme $\s$:
\be
\label{eq:beta_func_def}
    \beta_{\s}(\lambda_{\s}) \equiv \frac{\dd \, \lambda_{\s}(\mu)}{\dd \log(\mu^2)} \, .
\ee
This equation defines the running of the renormalized 't Hooft coupling $\lambda_{\s}(\mu)\equiv N g^2_{\s}(\mu)$ of the \SUN Yang--Mills theory and admits a perturbative expansion which is universal (i.e., scheme-independent) up to two loops:
\be
\label{eq:beta_func_pert}
    \beta_{\s}(\lambda_{\s}) \underset{\lambda_{\s} \,\to \, 0}{\sim} - \lambda_{\s}^2\left(b_0 + b_1 \lambda_{\s} + b_2^{(\s)} \lambda_{\s}^2 + \dots\right) \, ,
\ee
with $(4\pi)^2b_0=11/3$ and $(4\pi)^4b_1=34/3$. This renormalization group equation can be integrated exactly, defining a scheme-dependent, renormalization-group invariant $\Lambda$-parameter:
\begin{align}
    \label{eq:lambda_def}
    \frac{\Lambda_{\s}}{\mu} & = \left[b_0\lambda_{\s}(\mu)\right]^{-b_1/(2b_0^2)} \eee^{-1/[2b_0\lambda_{\s}(\mu)]} \,\eee^{-I_{\s}(\lambda_{\s}(\mu))} \, ,\\
    \label{eq:lambda_def2}
    I_{\s}(\lambda) & = \int_0^{\lambda} \dd x\,\left( \frac{1}{2\beta_{\s}(x)} + \frac{1}{2b_0x^2} - \frac{b_1}{2b_0^2x} \right) \, .
\end{align}
The values of $\Lambda$ in two schemes $\s,\s^\prime$ can be matched exactly through a one-loop calculation:
\be
    \label{eq:Lambda-conversion-generic}
    \log\left(\frac{\Lambda_{\s^\prime}}{\Lambda_{\s}}\right) = \frac{c_{\s\s^\prime}}{2b_0} \, ,
\ee
with
\be
    \lambda_{\s^\prime} = \lambda_{\s} \left(1 + c_{\s\s^\prime}\lambda_{\s} + \dots\right) \, .
\ee

By introducing two scales $\mu_1$ and $\mu_2$, one obtains the exact relation:
\be
    \label{eq:lambda_step_scal}
    \frac{\Lambda_{\s}}{\mu_1} = \frac{\Lambda_{\s}}{\mu_2} \exp\left\{-\int_{\lambda_{\s}(\mu_2)}^{\lambda_{\s}(\mu_1)} \frac{\dd x}{2\beta_{\s}(x)} \right\}\,.
\ee 
To determine $\Lambda_{\s}$ in terms of a hadronic scale, we take $\mu_1=\muhad$ in the non-perturbative regime and $\mu_2=\mupt$ in the perturbative regime. The scale $\muhad$ is defined such that both $\lambda_{\s}(\muhad)$ and $\muhad$ can be accurately determined from lattice simulations.

The key idea of the step-scaling method~\cite{Luscher:1991wu} is to determine the non-perturbative running of the coupling by relating its values at scales differing by a fixed factor, typically two. If the coupling $\lambda_{\s}(2\mu)$ can be obtained from $\lambda_{\s}(\mu)$, then by iterating this transformation $k$ times it is possible to reach a perturbative scale $\mupt = 2^k \muhad$. At this scale, $\Lambda_{\s}/\mupt$ can be obtained using perturbation theory,
\be\label{eq:Lambda_perturbative}
\begin{aligned}
        \left.\frac{\Lambda_{\s}}{\mupt}\right\vert_{\mathrm{pt}} = &\left[b_0\lambda_{\s}(\mupt)\right]^{-b_1/(2b_0^2)} \eee^{-\frac{1}{2b_0\lambda_{\s}(\mupt)}}\eee^{-I_{\s}^{(n)}[\lambda_{\s}(\mupt)]} \\
     & + \mathcal{O}[\lambda_{\s}^{n-1}(\mupt)] \, ,
\end{aligned}
\ee
where $I_{\s}^{(n)}(\lambda)$ is the perturbative truncation of $I_{\s}(\lambda)$ in \cref{eq:lambda_def2} evaluated with the $n$-loop $\beta$-function.
Instead, the non-perturbative running from $\muhad$ to $\mupt$ can be obtained by taking the continuum limit of lattice determinations of the coupling. Taking $k$ steps $\mu\to 2\mu$ from $\muhad$ to $\mupt$, the integral in the exponent of \cref{eq:lambda_step_scal} reduces to,
\be\label{eq:step_scaling_factor}
\begin{aligned}
    -\int_{\lambda_{\s}(\mupt)}^{\lambda_{\s}(\muhad)} \frac{\dd x}{2\beta_{\s}(x)} & = -\int_{\mupt}^{\muhad} \dd \log(\mu)\\
    & = \log\left(\frac{\mupt}{\muhad}\right) = \log\left(2^k\right) \, ,
\end{aligned}
\ee
leading to:
\be
    \label{eq:Lambda_with_step_scaling}
    \Lambda_{\s} = \lim_{\lambda_{\s}(\mupt) \,\to\, 0 } \left.\frac{\Lambda_{\s}}{\mupt}\right\vert_{\mathrm{pt}} 2^k \muhad\, .
\ee
In practice, thus, step scaling requires determining the running of the coupling over a wide range of scales, ensuring a controlled connection between the non-perturbative and perturbative regimes. The matching with perturbation theory can be performed for several sufficiently small couplings in order to ascertain the size of the perturbative corrections and to take a reliable $\lambda_{\s}(\mupt) \to 0$ limit.

It is convenient to introduce the so-called \emph{step-scaling function},
\be
    \label{eq:step-scaling-func-cont}
    \sigma_\s(u) = \lambda_{\s}(\mu/2)\Big\vert_{\lambda_{\s}(\mu) \,= \, u}\, ,
\ee
which describes how the coupling evolves under a discrete change of the renormalization scale $\mu$. Starting from $u_0 = \lambda_{\s}(\muhad)$, the sequence of couplings at higher energies is obtained iteratively as
\be
    \label{eq:step-scaling-iteration}
    u_{k+1} = \sigma_s^{-1}(u_k) \, .
\ee
In \cref{sec:results}, we will explain how to determine the inverse of the step-scaling function, $\sigma_s^{-1}(u)$, from our lattice simulations.

To run the scale $\mu$, it is convenient to work in a finite-volume renormalization scheme~\cite{Luscher:1991wu}, where the renormalization scale is set by the box size:
\be
    \label{eq:renorm-finite-vol}
    \mu = \frac{1}{c\ell} \, ,
\ee
with $\ell$ the length of the box and $c$ an arbitrary $\mathcal{O}(1)$ parameter that defines the scheme. For a symmetric lattice with size $L$ and spacing $a$, the physical size is expressed as $\ell = La$. 

\subsection{The Twisted Gradient Flow scheme}\label{sec:TGF-scheme}

As anticipated in \cref{sec:intro}, we implement the step-scaling method within the \emph{Twisted Gradient Flow} (TGF) scheme~\cite{Ramos:2014kla,Bribian:2019ybc,Bribian:2021cmg}. In this finite-volume setup, the \SUN Yang--Mills theory is defined on an asymmetric box of size $\ell^2\times\ell_s^2$, where the short extent is $\ell_s = \ell / N$. Periodic boundary conditions (\PBC) are imposed along the long space-time directions, $\mu=0,3$, while the $(1,2)$ plane (corresponding to the short directions) is endowed with twisted boundary conditions (\TBC)~\cite{tHooft:1979rtg}. The gauge fields therefore fulfill the following boundary conditions:
\be
A_\mu(x+\ell_\nu \hat{\nu}) = \Gamma_\nu A_\mu(x) \Gamma_\nu^\dagger,
\label{eq:TBC}
\ee
with $\Gamma_\mu\in$ \SUN satisfying,
\be
\Gamma_\mu \Gamma_\nu = z_{\mu\nu} \Gamma_\nu \Gamma_\mu, \quad z_{\mu\nu} = \eee^{\ii 2\pi \frac{ n_{\mu \nu}}{N}} \in \mathbb{Z}_N,
\label{eq:twist_eaters}
\ee
where $n_{12}=-n_{21}= k \ne 0 \text{ (mod } N)$ in the twisted plane, and $n_{\mu\nu} = 0 \text{ (mod } N)$ otherwise (the choice of $k$ will be discussed later).

This choice of geometry is motivated by the idea of \emph{twisted volume reduction}~\cite{PhysRevLett.48.1063,Gonzalez-Arroyo:1982hwr,Gonzalez-Arroyo:1982hyq,Gonzalez-Arroyo:2010omx,Gonzalez-Arroyo:2014dua} (see also Refs.~\cite{GarciaPerez:2014cmv,GarciaPerez:2020gnf} for reviews), a framework commonly used to study the large-$N$ limit of \SUN gauge theories~\cite{Gonzalez-Arroyo:2012euf,Gonzalez-Arroyo:2013bta,GarciaPerez:2014azn,GarciaPerez:2015rda,Gonzalez-Arroyo:2015bya,Perez:2020vbn,Butti:2022sgy,Butti:2023hfp,Bonanno:2023ypf,Bonanno:2024bqg,Bonanno:2024onr,Bonanno:2025hzr,Bonanno:2025bla,Hamada:2025whg}. As first established by Eguchi and Kawai~\cite{PhysRevLett.48.1063}, large-$N$ Yang--Mills theories exhibit a dynamical equivalence between color and space-time degrees of freedom, leading to volume independence for $N=\infty$. This property holds provided that center symmetry is unbroken, a condition that can be achieved via a suitable choice of twisted boundary conditions (for other approaches to implement large-$N$ volume reduction see~\cite{BHANOT198247,Gross:1982at,Kiskis:2002gr,Narayanan:2003fc,Kovtun:2007py,Unsal:2008ch,Neuberger:2020wpx,Yaffe:2026ouz}). While twisted volume reduction holds only in the large-$N$ limit, the adoption of \TBC at finite values of $N$ has the benefit of effectively enlarging the space-time volume. In particular, following perturbation theory, a system defined on a box with two short twisted directions and a volume $\ell^2\ell_s^2 = \ell^4 /N^2$ has finite-volume effects controlled primarily by an effective symmetric volume $\ell^4$. The use of \TBC also offers additional advantages. In perturbation theory, it allows for an expansion of observables in terms of integer powers of the coupling constant, unlike the case of periodic boundary conditions~\cite{Luscher:1982ma}. Moreover, on the lattice it avoids the $\mathcal{O}(a)$ effects present, for instance, in the Schr\"odinger Functional scheme~\cite{Luscher:1992an}.

The renormalized coupling of the TGF scheme is defined through the gradient flow~\cite{Narayanan:2006rf,Lohmayer:2011si,Luscher:2009eq}, a smoothing procedure that evolves the continuum gauge fields according to the equation
\be\label{eq:gradient_flow}
\hspace{-0.1\baselineskip}\partial_t B_\mu (x, t) = D_\nu F_{\nu \mu} (x, t), \quad B_\mu (x, t = 0) = A_\mu (x) \, ,
\ee
where $D_\mu$ and $ F_{\mu \nu}$ denote the covariant derivative and field-strength tensor at flow time $t$, while $A_\mu(x)$ stands for the bare gauge field. The flow suppresses UV fluctuations in the gauge configurations up to a smoothing radius $\sqrt{8t}$, and observables constructed with gauge fields at $t > 0$ are automatically renormalized at a renormalization scale given by $\mu^{-1} = \sqrt{8t}$~\cite{Luscher_2011}. 

As discussed in \cref{sec:strategy}, in finite-volume renormalization schemes the renormalization scale is linked to a fraction $c$ of the physical size $\ell$, $\mu^{-1} = c\ell$. Thus, the value of a dimensionless observable at flow time $t = (c\ell)^2/8$ can be used to define a renormalized coupling, which in the TGF scheme is
\be
    \label{eq:coupling_continuum}
    \lambdaTGF\left(\mu = \frac{1}{c \ell}\right) = \frac{1}{\mathcal{N}(c)} \frac{\braket{t^2 E(t)}}{N}\Big\vert_{\sqrt{8 t} \, = \, c \ell} \, \, ,
\ee
where $E(t)$ is the energy density evaluated on the flowed fields:
\be
    E(t) = \frac{1}{2} \Tr \left \{F_{\mu \nu} (x, t)F_{\mu \nu} (x, t)\right\}\, .
\ee
Note that, by virtue of twisted volume reduction, the long extent $\ell$ is taken as the representative length scale of the twisted box $\ell^2\times\ell_s^2$. The normalization factor $\mathcal{N}(c)$ is given by:
\begin{align}
\label{eq:flow_norm}
    \mathcal{N}(c) &= \frac{3\mathcal{A}(\pi c^2)}{128 \pi^2}, \\
    \nonumber\\[-1em]
    \mathcal{A}(x) &= x^2\theta_3^2(0,\ii x)\left[\theta_3^2(0,\ii x) - \theta_3^2(0,\ii xN^2)\right]\, ,
\end{align}
with $\theta_3(z,\ii x)=x^{-1/2}\sum_{m\in \mathbb{Z}}\exp[-\pi(m-z)^2/x]$ the Jacobi $\theta_3$ function. This ensures that, at lowest order of perturbation theory, $\lambdaTGF = \lambdaMS + \mathcal{O}(\lambdaMS^2)$. Each value of $c$, which can be chosen freely, defines a different scheme; here we adopt $c=0.3$.

The conversion factor between the $\Lambda$-parameters in the TGF and $\MStext$ schemes in \cref{eq:Lambda-conversion-generic} is known~\cite{Bribian:2019ybc},
\be\label{eq:Lambda-conversion}
\log\left(\frac{\LambdaTGF}{\LambdaMS}\right) = \frac{3}{22}\left[\frac{11}{3}\gamma_{_{\scriptscriptstyle{\rm E}}} + \frac{52}{9} - 3\log 3 + \mathcal{C}_1\right] \, ,
\ee
with $\gamma_{_{\scriptscriptstyle{\rm E}}}\simeq 0.5772\dots$ the Euler--Mascheroni constant, and where $\mathcal{C}_1$ has been calculated numerically for several combinations of $N$ and $c$, including $N=3,5,8$ at $c=0.3$ and the choice of twist necessary for this work. Using the same methodology of~\cite{Bribian:2019ybc}, we found $\mathcal{C}_1=0.508(4)$, $0.597(14)$, and $0.615(14)$ for $N=3,5,8$ and $k=1,2,3$ respectively (see next section for more details about the choices of \TBC adopted in this study). Thus, the TGF can be used as an intermediate scheme to determine $\LambdaMS$. However, only the two universal coefficients of the beta function $\beta_{\TGF}$ are known, and therefore very high energy scales must be reached with the step-scaling procedure in order to reliably match it with perturbation theory using \cref{eq:Lambda_perturbative}.

\section{Lattice setup}\label{sec:lattice}

We discretize the four-dimensional pure-gauge \SUN theories for $N=3,5,8$ using the Wilson plaquette action on asymmetric lattices with a short size $L_s$ along the two directions $\mu=1,2$ and a long size $L = NL_s$ along $\mu=0,3$. In physical units, the lattice volume is $\ell^2\times\ell_s^2$ with $\ell = aL$, $\ell_s = aL_s$, and $a$ the lattice spacing. We impose \PBC along the long directions and \TBC~\cite{tHooft:1979rtg} in the plane of the short directions. The lattice action is given by:
\be\label{eq:lattice_action_TBC}
S_{\scriptscriptstyle{\rm W}}[U] = -N b\sum_{x,\mu \, \neq \, \nu} Z_{\mu\nu}^*(x)\,  \Tr \left[ P_{\mu\nu}(x)\right],
\ee
where $b = 1/\lambda_0 = \beta_g/(2N^2)$ is the inverse bare 't Hooft coupling ($\beta_g$ is the usual inverse lattice bare coupling), and $P_{\mu\nu}(x)$ is the plaquette,
\be
P_{\mu\nu}(x) = U_\mu(x) U_\nu(x+a\hat{\mu}) U_\mu^{\dag}(x+a\hat{\nu}) U^{\dag}_\nu(x) \, .
\ee
The factor $Z_{\mu\nu}(x)$ implements \TBC. For $\mu < \nu$,
\be\label{eq:twist_factor}
\begin{split}
Z_{\mu\nu}(x) = Z_{\nu\mu}^*(x) =
\begin{cases}
\eee^{\ii \frac{2 \pi k}{N}}, & \makecell{(\mu,\nu)=(1,2)\\ x_{\mu}=x_\nu=0} \,, \\
\\[-1em]
1, & \text{otherwise,}
\end{cases}
\end{split}
\ee
where $k$ is an integer coprime with $N$. As in Ref.~\cite{Bonanno:2024nba}, $k$ and $N$ are chosen such that they are separated by two steps in the Fibonacci sequence, that is, $k=1,2,3$ for $N=3,5,8$, respectively. This selection is motivated by previous studies of twisted boundary conditions in the large-$N$ limit~\cite{Gonzalez-Arroyo:2010omx,GarciaPerez:2013idu,GarciaPerez:2014cmv,Chamizo:2016msz,Perez:2017jyq,GarciaPerez:2018fkj,Bribian:2019ybc}, where it has been shown that the choice of $k$ is crucial to ensure volume reduction and a smooth approach to the large-$N$ limit. Selecting $k$ and $N$ as next-to-consecutive terms of the Fibonacci sequence $F_n$, i.e., $k = F_{n-2}$ and $N = F_n$, saturates the bound for the absence of tachyonic instabilities and provides a controlled definition of the twisted theory at large $N$~\cite{Chamizo:2016msz}.

The gauge configurations were generated with a combination of Cabibbo--Marinari~\cite{Cabibbo:1982zn} heatbath (HB)~\cite{Creutz:1980zw,Kennedy:1985nu} and over-relaxation (OR)~\cite{Creutz:1987xi} lattice sweeps. For most ensembles considered in this work, each update cycle consisted of one HB sweep followed by $n_{\ov} = L$ OR sweeps, and consecutive measurements were separated by $\Delta_{\meas} = 10$ update cycles. For the production of the most computationally demanding ensembles, specifically those with $L=64$ and $L=96$ for $N=8$, we tested different updating schemes to improve the efficiency of the Monte Carlo evolution. We found that reducing the number $n_{\ov}$ of OR sweeps per HB sweep as the inverse bare coupling $b$ gets smaller, down to $n_{\ov} = 4$ for $b \simeq 0.44$, leads to shorter autocorrelation times. The interval $\Delta_{\meas}$ between measurements was changed according to the autocorrelation time in order to have well-decorrelated samples. Independent sets of measurements obtained with different updating schemes were binned separately and combined only at the averaging stage. More precisely, the integrated autocorrelation time is estimated independently for each Monte Carlo run using the standard $\Gamma$-method~\cite{Madras:1988ei,Wolff:2003sm,Schaefer:2010hu}. This allows us to combine runs generated with different updating schemes, using the autocorrelation time to perform an initial binning of the raw data. Then, the binned data from all runs is combined and further divided into 20 jackknife bins, with bin boundaries chosen to lie within individual runs. Every subsequent calculation is carried out independently for each bin, ensuring that statistical fluctuations are propagated consistently through the entire pipeline of the analysis. The bins are recombined only at the final stage: the central value is obtained as their mean, while the statistical uncertainty is estimated from the variance of the 20 jackknife estimates.

Let us now discuss the relevant lattice observables. The flow equation in \cref{eq:gradient_flow} is discretized as,
\be
\label{eq:wilson_flow}
\begin{aligned}
\partial_t V_\mu(x, t) & = -\frac{1}{Nb}\left(\partial_{x,\mu}S_{\scriptscriptstyle{\rm W}}[V]\right)V_\mu(x, t) \, ,\\ V_\mu(x, 0) & = U_\mu(x) \, ,
\end{aligned}
\ee
where $\partial_{x,\mu}S_{\scriptscriptstyle{\rm W}}[V]$ is the Lie-algebra valued derivative of the Wilson action. We integrate the so-called Wilson flow in \cref{eq:wilson_flow} with the adaptive third-order Runge--Kutta method described in Ref.~\cite{Fritzsch:2013je}.

To discretize the energy density, entering the definition of the renormalized coupling in \cref{eq:coupling_continuum}, we adopt the \emph{clover} discretization:
\be \label{eq:Eclover}
a^4E_{\clov}(t) = -\frac{1}{2} \Tr\left[C_{\mu\nu}(x,t)C_{\mu\nu}(x,t)\right] \, ,
\ee
where $C_{\mu\nu}(x,t)$ is the clover operator on the $(\mu,\nu)$ plane at the site $x$, evaluated after the gauge links have been evolved for a flow time $t$. It is defined as the anti-Hermitian and traceless part of:
\be
\begin{aligned}
	\frac{1}{4}\, \bigg[& Z^*_{\mu\nu}(x) P_{\mu\nu}(x,t) + Z^*_{\mu\nu}(x-a\hat{\nu}) P_{-\nu\mu}(x,t)  \\   
	&+ Z^*_{\mu\nu}(x-a\hat{\mu}) P_{\nu-\mu}(x,t)  \\   
	&+ Z^*_{\mu\nu}(x-a\hat{\mu}-a\hat{\nu}) P_{-\mu-\nu}(x,t) \bigg],
\end{aligned}
\ee
where $U_{-\mu}(x, t) = U_{\mu}^{\dagger}(x-a\hat{\mu},t)$. Then, the coupling is defined as~\cite{Fritzsch:2013yxa}:
\be
    \label{eq:coupling_proj}
    \lambdazeroTGF \left(\mu = \frac{1}{c \ell}\right) = \frac{1}{\mathcal{N}_{\scriptscriptstyle{\rm L}}(c,L)}
    \left.\frac{\braket{t^2 E_{\clov}(t)\delta_{Q,0}}}{N\braket{\delta_{Q,0}}}\right\vert_{\sqrt{8 t} \, = \, c \ell} \, .
\ee
Here, the lattice normalization $\mathcal{N}_{\scriptscriptstyle{\rm L}}(c,L)$ is chosen so as to eliminate the leading-order lattice artifacts in perturbation theory: 
\begin{equation*}
\mathcal{N}_{\scriptscriptstyle{\rm L}}(c,L) = \frac{c^4}{128}\sum_{\mu\,\neq\,\nu} \sum_{q}^{\prime} \eee^{-\frac{1}{4}c^2L^2\hat{q}^2}\frac{\sin^2(q_\nu)}{\hat{q}^2}
\cos^2\left(\frac{q_\mu}{2}\right),
\end{equation*}
where $\hat q_\mu \equiv 2\sin(q_\mu/2)$ stands for the lattice momentum --- with $q_\mu = 2 \pi n_\mu /L$ and $n_\mu = 0,\,1, \, \dots,\, L-1$ --- and where the prime in the sum denotes the exclusion of momenta with both components in the twisted plane satisfying $L \,q_i \propto 2 N \pi$.

Another important ingredient entering the lattice definition of the coupling in \cref{eq:coupling_proj} is the $Q=0$ \emph{topological projection}, achieved via the quantity $\delta_{Q,0}$ (to be specified below) in \cref{eq:coupling_proj}. The topological projection was introduced in Ref.~\cite{Fritzsch:2013yxa} as a suitable choice of lattice renormalization scheme for step-scaling studies, becoming advantageous when global topological fluctuations suffer from long autocorrelation times at fine lattice spacings, a problem known as topological freezing~\cite{Alles:1996vn,DelDebbio:2004xh,Schaefer:2010hu}. The results of Ref.~\cite{Bonanno:2024nba} lent additional support to this procedure, since it was shown that a non-projected definition of $\lambda_{\TGF}$ (supplemented with an algorithm that avoids topological freezing) leads to consistent results for the determination of the $\Lambda$-parameter (see also~\cite{Luscher:2014kea,Albandea:2021lvl} on this point). In order to implement the projection, we define the lattice topological charge from the flowed clover discretization:
\be
Q_{\clov}(t) = \sum_{x,\,\mu\nu\rho\sigma}\frac{\varepsilon_{\mu\nu\rho\sigma}}{32\pi^2}\Tr\left[C_{\mu\nu}(x,t)C_{\rho\sigma}(x,t)\right] \, ,
\ee
and assign to each configuration the integer charge
\be\label{eq:topcharge_lat_def}
Q = \mathrm{round}\left[Q_{\clov}\left(t = \frac{c^2\ell^2}{8}\right)\right]\, .
\ee
We verified that $Q_{\clov}(t)$ reaches a near-integer plateau in $t$ before $t = (c\ell)^2/8$, thus it can be safely rounded to the closest integer at the flow time defining the coupling. Therefore, we define $\delta_{Q,0}=1$ for gauge configurations with $Q=0$, and $\delta_{Q,0}=0$ otherwise.

\begin{figure*}[!t]
\centering
\includegraphics[scale=0.33]{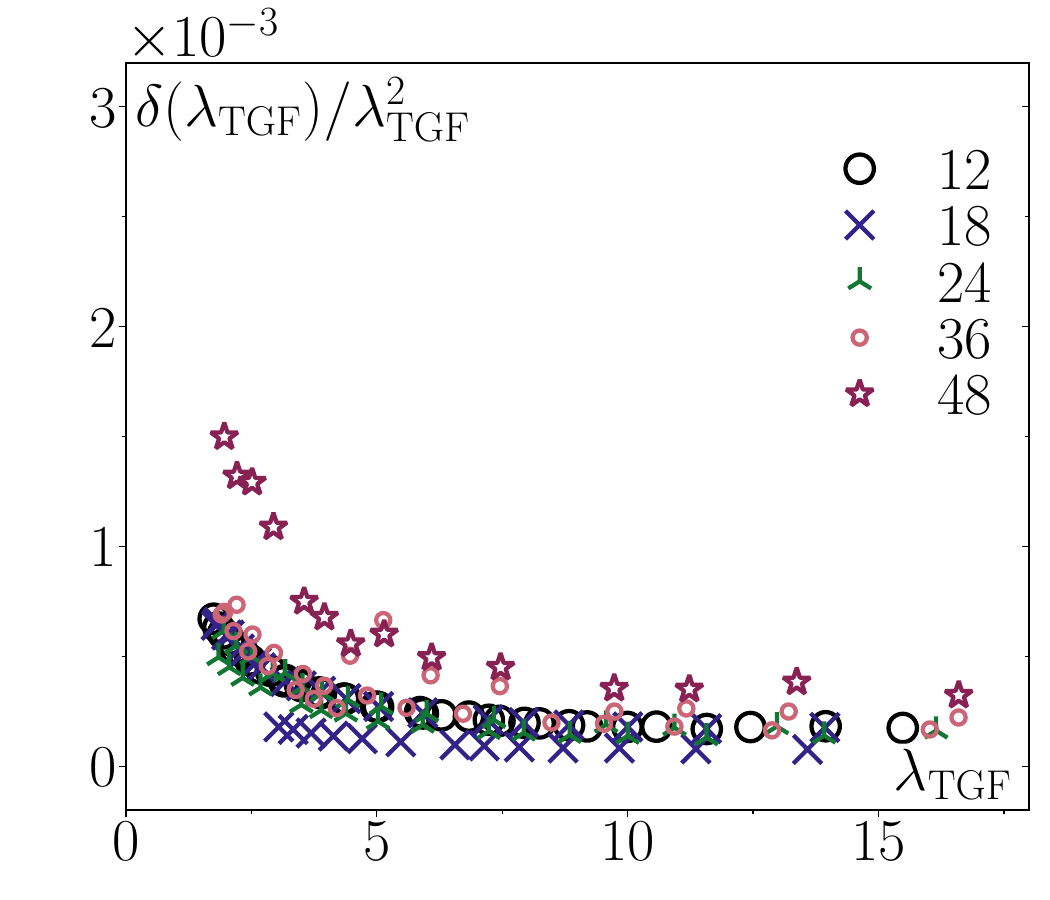}
\includegraphics[scale=0.33]{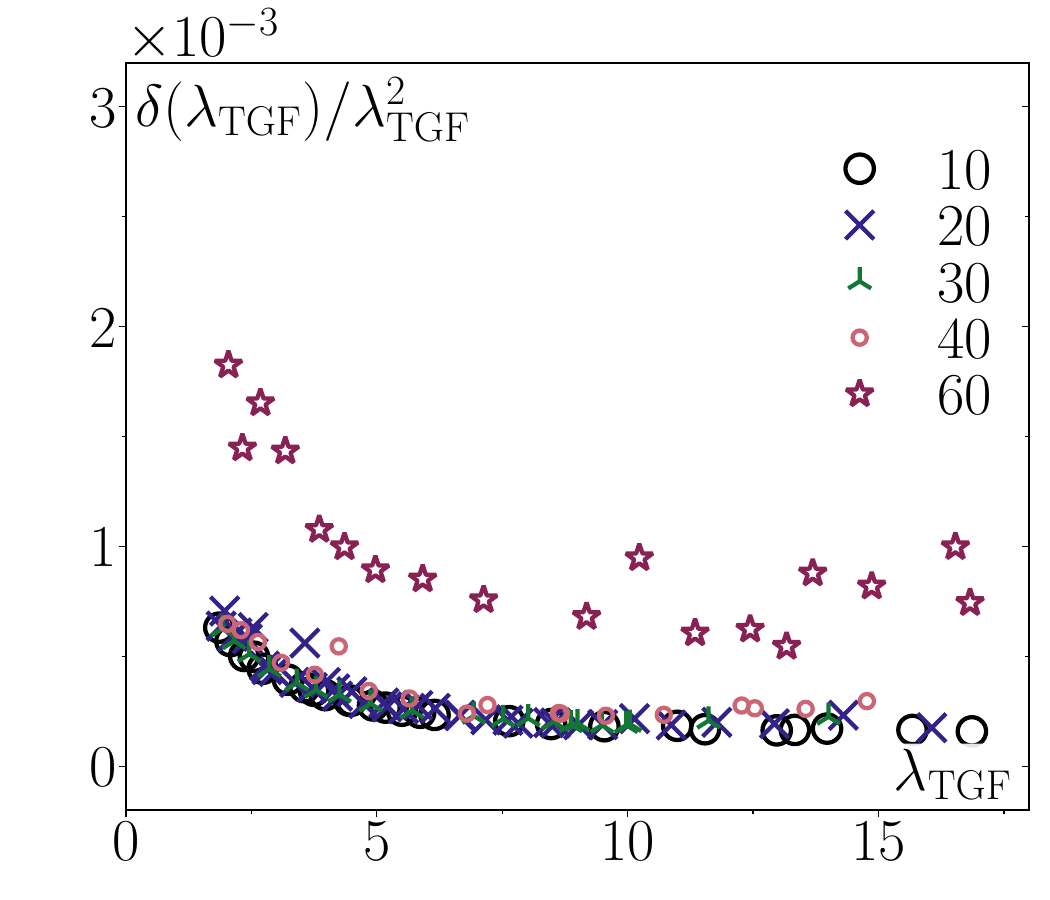}
\includegraphics[scale=0.33]{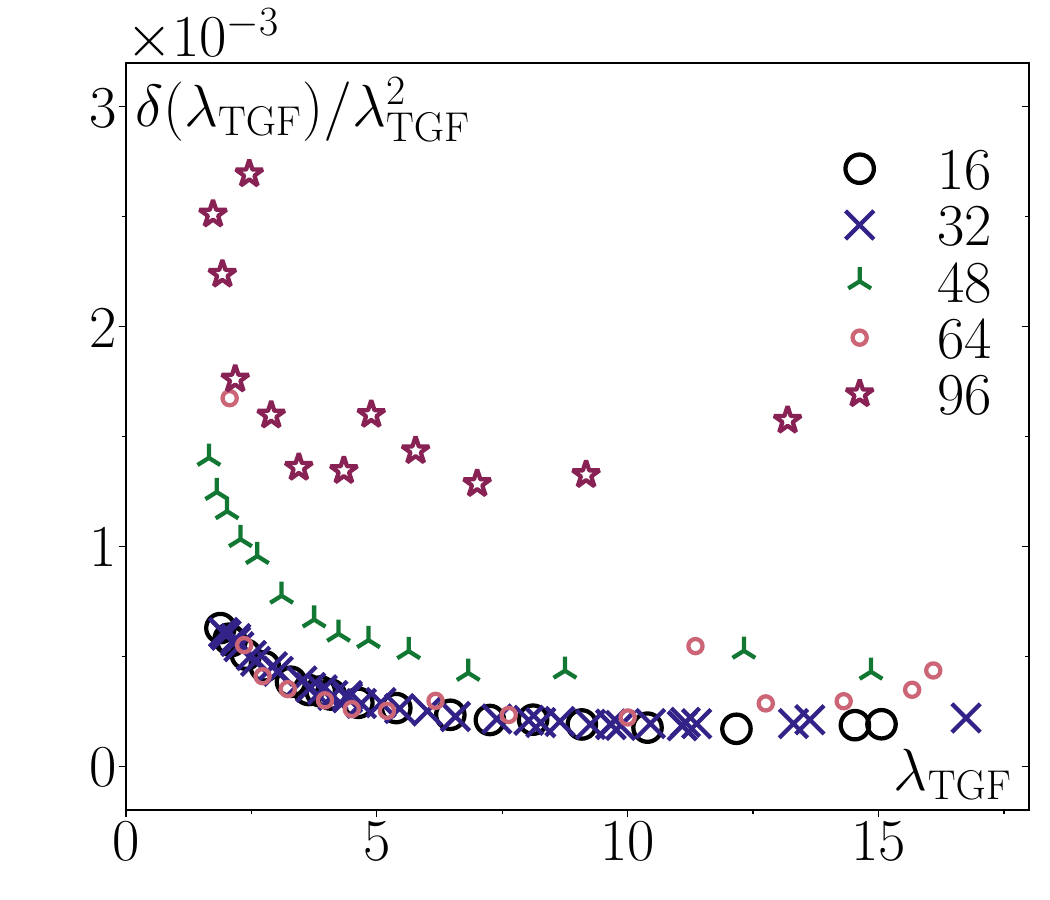}
\caption{Relative accuracy of the renormalized 't Hooft coupling determination. From left to right: $N=3,5,8$.}
\label{fig:lambda_err}
\end{figure*}

The full list of simulation points, statistics, and the measured values of the renormalized coupling for \SU{5} and \SU{8} is reported in \cref{tab:rawdata} in Appendix \ref{appendix:rawdata}. The full \SU{3} dataset can instead be found in the previous studies~\cite{Bribian:2021cmg,Bonanno:2024nba}. The statistical precision was chosen with the aim of keeping the uncertainty of the inverse coupling approximately constant over the range of simulated couplings. Equivalently, the quantity $\delta\lambda_{\TGF}/\lambda_{\TGF}^2$ is kept approximately uniform. The resulting uncertainties are shown in \cref{fig:lambda_err}. In practice, the largest lattices and the weakest-coupling points are the most computationally demanding, and their statistical errors dominate the final uncertainty.

\section{Lattice methods and numerical results}\label{sec:results}

\subsection{Step scaling on the lattice}\label{subsec:step_scaling_lattice}

In this section, we discuss the practical lattice implementation of the step-scaling method described in \cref{sec:step-scaling} for determining the $\Lambda$-parameter through the non-perturbative running of the TGF coupling. Since the couplings and the corresponding step-scaling functions are determined only in the TGF scheme, we omit the subscript TGF until the conversion from $\LambdaTGF$ to $\LambdaMS$.

First, we explain how the simulation points were chosen. To extract a continuum limit, every energy scale $\mu = 1/(caL)$ in the step-scaling sequence must be simulated at different lattice spacings. To control lattice artifacts, it is sufficient to have $1/L^2 \propto (a\mu)^2 \ll 1$, since $\mu$ is the only scale that needs to be resolved. This is how step scaling avoids the window problem of asymptotic scaling mentioned in \cref{sec:intro}. As described in \cref{sec:lattice}, in the TGF scheme the lattice has two short directions $L_s = L / N$, so $L$ must also be a multiple of $N$.

The next step is to choose the starting low-energy scale $\muhad$ of the step-scaling sequence in \cref{eq:step-scaling-iteration}. On the lattice, this is implicitly set by choosing the starting renormalized coupling,
\be
u_0 = \lambda(\muhad) \, .
\ee
The step-scaling procedure only gives the $\Lambda$-parameter in units of $\muhad$, as it can be seen from \cref{eq:Lambda_with_step_scaling}. For a meaningful result, $\muhad$ must then be determined in units of a common reference scale, which we chose to be the gradient flow scale $t_0$. This requires an independent set of infinite-volume simulations dedicated to scale setting, which put a constraint on the choice of $u_0$. In particular, the conversion factor $\muhad\sqrt{8t_0}$ from $\Lambda/\muhad$ to $\Lambda\sqrt{8t_0}$
is determined as
\be\label{eq:muhadsqrt8t0_1}
\muhad\sqrt{8t_0} = \lim_{a \, \to \, 0} \Big (a\muhad \times \sqrt{\frac{8t_0}{a^2}} \Big )\, .
\ee
The scale $\sqrt{8t_0}/a$ must be determined at a set of lattice spacings $a \ll \muhad^{-1}$. At the same time, controlling the finite-volume effects on $t_0$ requires physical sizes $\ell \gtrsim 3\sqrt{8t_0}$. Therefore, scale setting is computationally feasible only if $\muhad\sqrt{8t_0} = \mathcal{O}(1)$, so that the required lattice sizes $\ell / a \gg \muhad\sqrt{8t_0}$ are not too large. The necessary scale setting simulations for the present study were presented in the dedicated paper~\cite{Bonanno:2025kfd}, where we determined $t_0$ and other gradient-flow scales for the bare couplings required to compute $\muhad$. These scale-setting results will be discussed later in \cref{subsec:muhad_sqrt8t0}, where they will be used to convert $\LambdaMS/\muhad$ to $\LambdaMS\sqrt{8t_0}$.

Now, let us discuss in more detail how step scaling can be implemented on the lattice. For a given value of $N$, let the couplings measured on two lattices with sizes $L$ and $2L$ be:
\be
\begin{aligned}
u &\equiv \lambda(b,L)\, , \\
\Sigma(u,L) &\equiv \lambda(b,2L) \, .
\end{aligned}
\ee
If the renormalization scale of the coupling measured on the small lattice is $\mu = 1/(caL)$, the one of the large lattice is $\mu/2$, having kept $b$ and thus $a$ fixed. So, $\Sigma(u,L)$ represents a determination of the step-scaling function $\sigma(u)$ at finite lattice spacing. Equivalently, the two renormalized couplings, which we now denote as
\be
\begin{aligned}
\label{eq:step-scaling-func-latt}
\sigma &\equiv \lambda(b,2L)\, , \\
{\cal U}(\sigma,L) &\equiv \lambda(b,L) \, ,
\end{aligned}
\ee
constitute a determination of the inverse of $\Sigma(u,L)$, that is, a lattice measurement of the inverse step-scaling function $u(\sigma)$. With this in mind, we consider two possibilities to implement step scaling, that is, to construct the sequence of renormalized couplings $u_{k+1} = u(u_{k})$ starting from $u_0$.

In one approach, lattice results for $\lambda(b, 2L)$ are parametrized and fitted as a function of $b$ at each simulated size $2L$. The fits can be used to interpolate the values $b(u_k, 2L)$ of $b$ giving a line of constant physics (LCP) determined by $\lambda(b, 2L) = u_k$. Then, a similar fit and interpolation of $\lambda(b, L)$ is used to determine $u_{k+1}$ as
\be
u_{k+1} = \lim_{1/L \, \to \, 0} \lambda\left(b(u_k, 2L), L\right)\, .
\ee
This continuum limit is needed because halving the lattice sizes of the LCP for $u_k$ gives an LCP for $u_{k+1}$ only up to lattice artifacts.

In the other approach, we parameterize the inverse step-scaling function on the lattice, ${\cal U}(\sigma,L)$, and its lattice artifacts in terms of $\sigma$ and $L$ in order to obtain the continuum function $u(\sigma)$ from a global fit of all the results for $\lambda(b, L)$ and $\lambda(b, 2L)$. Then, the results of the fit can be used to construct the step-scaling sequence by repeated evaluation of $u(\sigma)$ as in \cref{eq:step-scaling-iteration}.

\begin{figure*}[!t]
\centering
\includegraphics[scale=0.33]{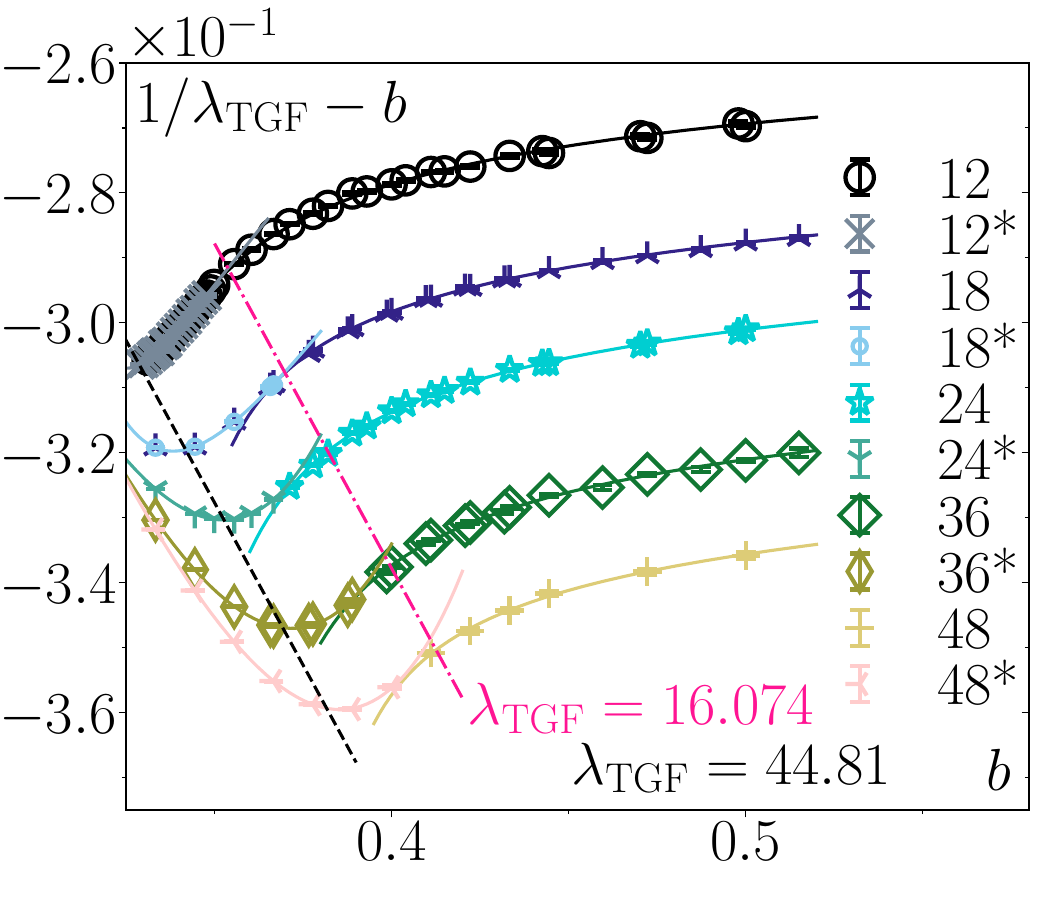}
\includegraphics[scale=0.33]{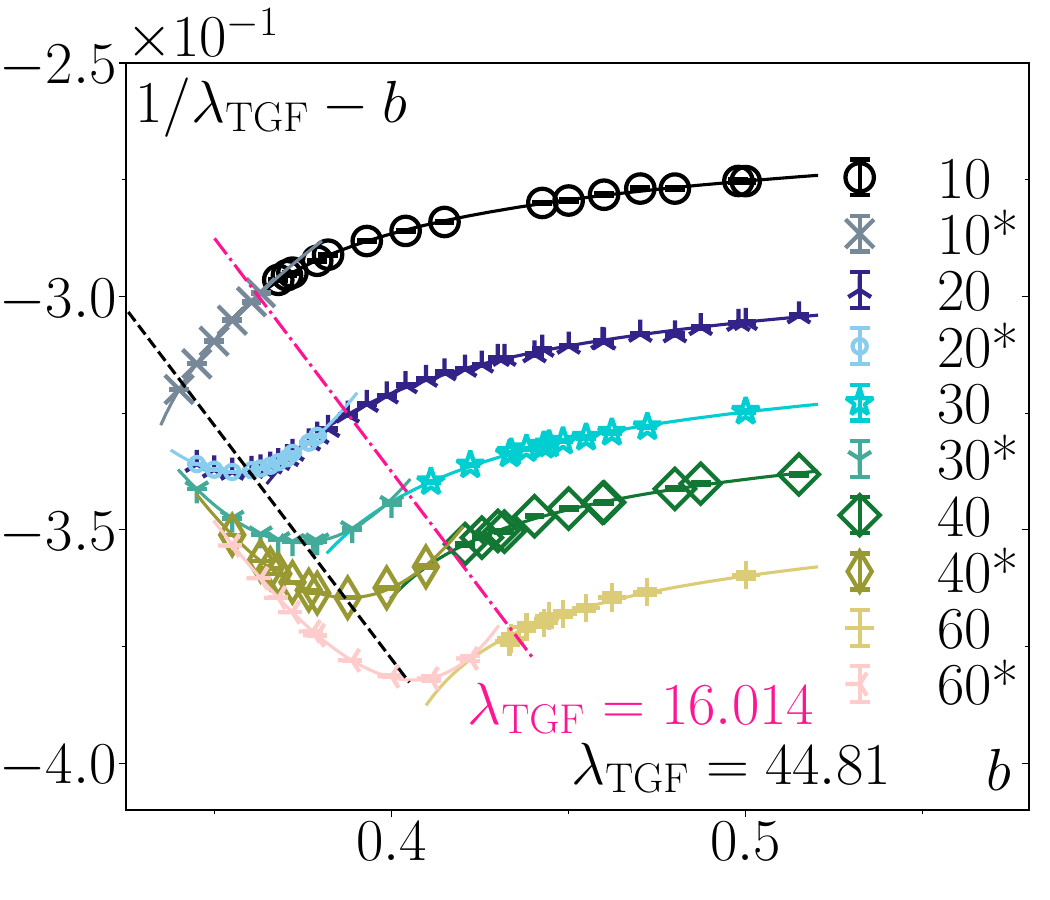}
\includegraphics[scale=0.33]{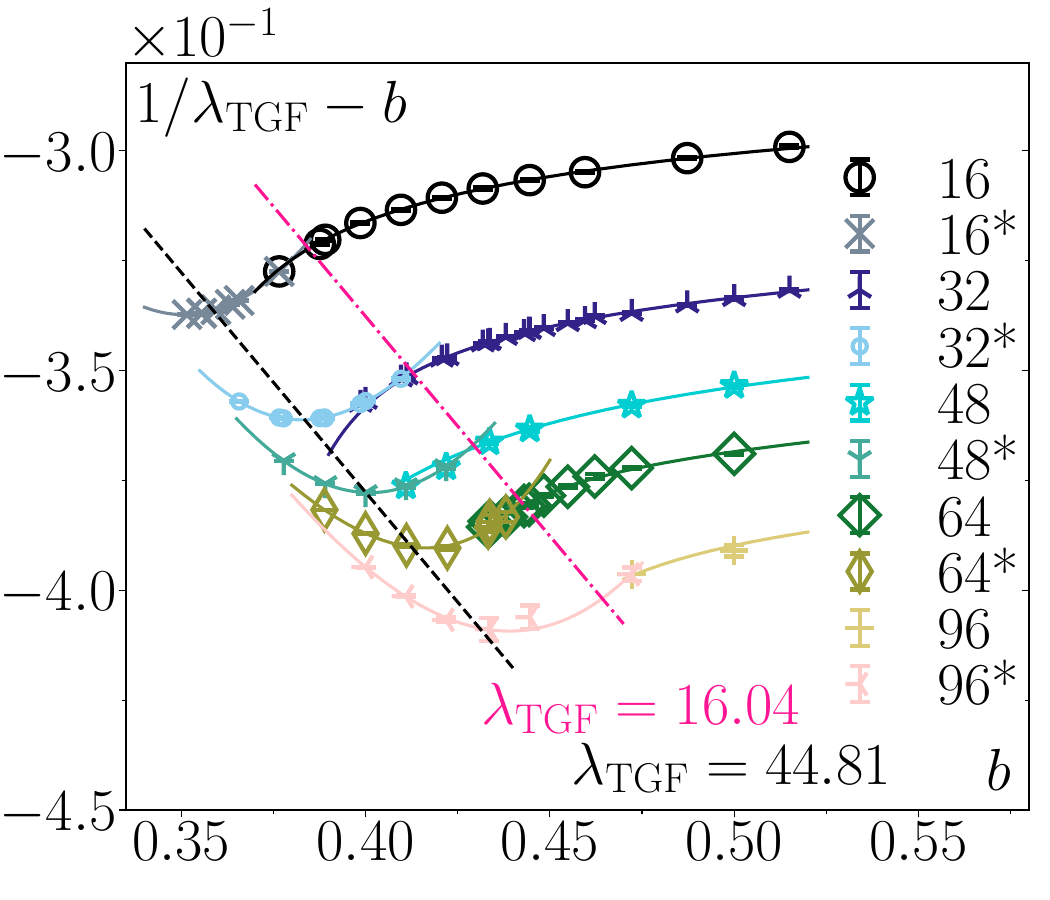}
\caption{Behavior of $1/\lambda - b$ as a function of $b$. The solid curves represent Padé fits of the form given in \cref{eq:lambda_vs_b}. The solid curves, with or without a $*$, correspond to fits performed over different ranges of $b$, motivated by the change in the behavior of the coupling around $b \simeq 0.40-0.45$. The dashed and dot-dashed lines correspond to lines of constant physics with fixed values of the renormalized coupling. From left to right: $N=3,5,8$.}
\label{fig:lambda_vs_b}
\end{figure*}

As will be shown in \cref{subsec:initial}, it is not possible to employ a single simple parameterization of $u(\sigma)$ across the entire relevant range of energies, since $\lambda(\mu)$ changes behavior for $\mu \lesssim 2\muhad$. Starting the step-scaling sequence directly at $2\muhad$ would make scale setting unfeasible by requiring lattice spacings that are a factor of two smaller. We therefore rely on both approaches. The first approach is used only for the initial step of the sequence, connecting $\muhad$ to $2\muhad$ by determining $u_1$ from the chosen $u_0$. The second approach is then used to run the coupling from $u_1$ towards the perturbative regime.

To summarize, our strategy is the following:
\begin{enumerate}
    \item Determine $u_1=\lambda(2\muhad)$, with $\muhad$ defined by the chosen $u_0=\lambda(\muhad)$, by fitting the $b$-dependence of $\lambda$ on the lattice. This is discussed in \cref{subsec:initial}.
    \item From a global fit of the values of $\lambda$ on the lattice, determine the inverse step-scaling function $u(\sigma)$ in a range of couplings $\sigma$ that includes $u_1$ and extends to the perturbative regime. This is discussed in \cref{subsec:step_scaling}.
    \item Determine $\LambdaMS/\muhad$ by using $u(\sigma)$ to run the coupling to sufficiently high energies in order to match with perturbation theory. This is discussed in \cref{subsec:LambdaTGF}.
    \item Convert $\LambdaMS/\muhad$ to $\LambdaMS\sqrt{8t_0}$ using the results of scale setting to determine $\muhad\sqrt{8t_0}$. This is discussed in \cref{subsec:muhad_sqrt8t0}.
\end{enumerate}

\subsection{Initial step from \texorpdfstring{$\muhad$}{muhad} to \texorpdfstring{$2\muhad$}{2 muhad}}\label{subsec:initial}

In this section, we describe the first step of the step-scaling sequence, that is, the choice of $u_0 = \lambdaTGF(\muhad)$ and the determination of $u_1 = \lambdaTGF(2\muhad)$. As anticipated in \cref{subsec:step_scaling_lattice}, this requires a dedicated approach because $u_0$ will turn out to be too deep in the strong-coupling region to use the same parameterization of the step-scaling function that we will employ from $u_1$ towards the perturbative regime.

The idea is to explicitly find the bare couplings $b_0 = b(u_0, 2L)$ that give an LCP at the chosen renormalized coupling $u_0$ with a set of large lattice sizes $2L$,
\be\label{eq:u0-tuning}
u_0 = \lambda(b_0, 2L) \, ,
\ee
which, recalling \cref{eq:renorm-finite-vol}, implicitly defines the renormalization scale $\muhad$ as
\be\label{eq:muhad_lattice}
\muhad^{-1} = c \times a(b_0) \times 2L \, .
\ee
Then, by definition, $u_1$ is the coupling at renormalization scale $2\muhad$, which means half the physical size. From \cref{eq:muhad_lattice}, $u_1$ can be obtained on the lattice from $\lambda(b_0, L)$, that is, the renormalized coupling at the same lattice spacings and half the lattice sizes, but only up to lattice artifacts. These can be removed by taking the continuum limit, given by $1/L \propto (a\muhad) \to 0$ being $\muhad$ fixed. Thus,
\be\label{eq:u1-limit}
u_1 = \lim_{1/L \, \to \, 0} \lambda\left(b_0, L\right)\, .
\ee

In principle, this initial step could be implemented by manually tuning the bare couplings $b_0$ on lattices of size $2L$ and then repeating the simulations with the same values of $b_0$ on lattices of size $L$. As a simpler alternative route, we instead simulated several values of $b$ in the strong-coupling region with lattice sizes $L$ and $2L$ in order to fit the measurements of $\lambda(b, L)$ and $\lambda(b, 2L)$ as functions of the bare coupling $b$. Then, we obtained the values of $b_0$ by using the fitted $\lambda(b, 2L)$ to solve \cref{eq:u0-tuning}, and the values of $\lambda(b_0, L)$ by interpolating the fitted $\lambda(b, L)$. One advantage of this approach is that it allows us to choose an optimal $u_0$ based on the resulting $u_1$ \emph{a posteriori} and without further simulations, as discussed below.

The parameters and results of the strong-coupling simulations dedicated to this step are also reported in \cref{tab:rawdata} in Appendix~\ref{appendix:rawdata}. Each combination of $N$ and $L$ was fitted independently as a function of $b$ with a Padé approximant,
\be
b \, \lambda(b, L) = \frac{a_0 + a_1 b + b^2}{a_2 + a_3 b + b^2} \, ,
\label{eq:lambda_vs_b}
\ee
which provides a stable parameterization over the explored range of couplings. The fits, shown in \cref{fig:lambda_vs_b}, are performed on the quantity $1/\lambda - b$, and yield values of $\chi^2$ per degree of freedom in the range $0.37$ to $1.98$. Two different Padé fits are performed: one in the weak-to-intermediate coupling region, covered by the continuum step-scaling sequence described in \cref{subsec:step_scaling}, and one in the strong-coupling region.

\begin{figure}[!t]
\centering
\includegraphics[scale=0.45]{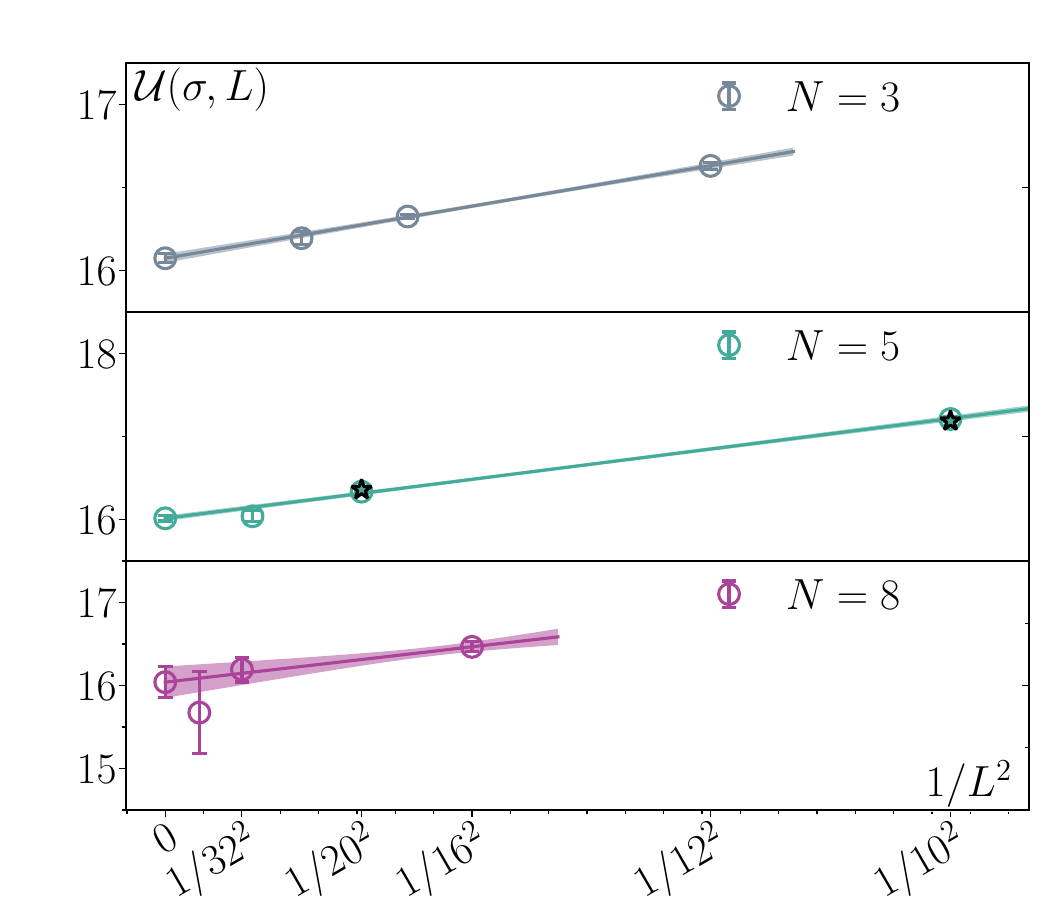}
\caption{Continuum extrapolations of the first coupling $u_1$ in the step-scaling sequence after the chosen $u_0 = 44.81$, as described in the main text. All data are reported in \cref{tab:u0}. The star markers in the central $N=5$ panel correspond to direct simulation, providing a cross-check of our Padé interpolations.}
\label{fig:initial_step}
\end{figure}

We can now discuss in more detail how $u_0$ was chosen for each value of $N$. As discussed in \cref{subsec:step_scaling_lattice}, one would like to choose $u_0$ in such a way that scale setting is feasible for the bare couplings $b_0$. This bounds feasible choices of $u_0$ from \emph{below}. At the same time, one would like $u_1$ to fall within the regime where a single parameterization of the step-scaling function can be employed up to perturbatively high energies. This bounds feasible choices of $u_0$ from \emph{above}. A reasonable choice that satisfies these two criteria for all values of $N$ is:
\begin{align}
 u_0  &= 44.81\,. 
\end{align}
This value is represented by the sloped black dotted lines in \cref{fig:lambda_vs_b}. Once this is fixed, the bare couplings corresponding to $\lambdaTGF (b, 2L)=u_0$ for each of the lattices of size $2L$ are determined using the Padé fits, cf.~\cref{tab:u0}. Finally, at the selected values of $b$, the Padé interpolation is again used to  determine $\mathcal{U}(u_0,L) \equiv \lambdaTGF(b,L)$, also reported in \cref{tab:u0}. The continuum-limit value of $u_1$ is obtained through a linear extrapolation of $\mathcal{U}(u_0,L)$ in $1/L^2$, as shown in \cref{fig:initial_step}, cf.~\cref{eq:u1-limit}. As a cross-check of this procedure, in two \SU{5} cases we compared the couplings extracted from the Padé fits with the results from actual simulations at the corresponding values of $b$ and $L$, finding perfect agreement, cf.~the star points in the central panel of \cref{fig:initial_step}. The resulting values of $u_1$ in the continuum limit are given by:
\begin{align}
 u_1  &= 16.074(22) \text{ for \SU{3}} \,,\\
 u_1  &= 16.014(39) \text{ for \SU{5}} \,,\\
 u_1  &= 16.04(21) \,\,\,\text{ for \SU{8}}\,, 
\end{align}
where errors have been estimated via a binned jackknife analysis, see \cref{sec:lattice}. They are represented in \cref{fig:lambda_vs_b} by the sloped dotted pink lines. The reduced $\chi^2$ values are $0.29, 3.45, 0.80$ for $N=3,5,8$, respectively. Since the fit quality of $N=5$ is only marginally acceptable ($p$-value $\simeq 6.3\%$), we also performed a further stability check in this case by fitting $1/{\cal U}$ linearly in $1/L^2$, finding a perfectly compatible result: $u_1=16.034(32)$.

\begin{table}[!t]
\centering
\begin{tabular}{ccccc}
\midrule
$N$ & $L$ & $b(u_0)$ & $\lambdaTGF^{\text{Padé}}(b,L)$ & $\lambdaTGF^{\text{Padé}}(b,2L)$  \\
\midrule
3 & 12 & 0.352668(38)  & 16.630(19)  & \multirow{4}{*}{44.81} \\
3 & 18 & 0.369260(33)  & 16.325(15)  &                        \\
3 & 24 & 0.381660(98)  & 16.194(40)  &                        \\
3 & $\infty$ & $\infty$& 16.074(22)  &                        \\
\addlinespace[0.5em]
5 & 10 & 0.359706(30)  & 17.2106(34) & \multirow{4}{*}{44.81} \\
5 & 20 & 0.386833(68)  & 16.335(26)  &                        \\
5 & 30 &  0.40446(18)  & 16.038(65)  &                        \\
5 & $\infty$ & $\infty$& 16.014(39)  &                        \\
\addlinespace[0.5em]
8 & 16 & 0.383526(34)  & 16.465(60)  & \multirow{4}{*}{44.81} \\ 
8 & 32 & 0.41231(32)   & 16.19(15)   &                        \\ 
8 & 48 & 0.4309(16)    & 15.67(50)   &                        \\ 
8 & $\infty$ & $\infty$&  16.04(21)  &                        \\
\bottomrule
\end{tabular}
\caption{Lattice measurements and interpolations of $\lambda$ entering the determination of $u_1 = \lambdaTGF(2\muhad)$, the first coupling of the step-scaling sequence after the chosen $u_0 = \lambdaTGF(\muhad)$. The continuum extrapolation of $u_1$ is shown in \cref{fig:initial_step}.}
\label{tab:u0}
\end{table}

\subsection{Continuum step-scaling function from a global fit of lattice data}\label{subsec:step_scaling}

\begin{figure*}[!t]
\centering
\includegraphics[scale=0.33]{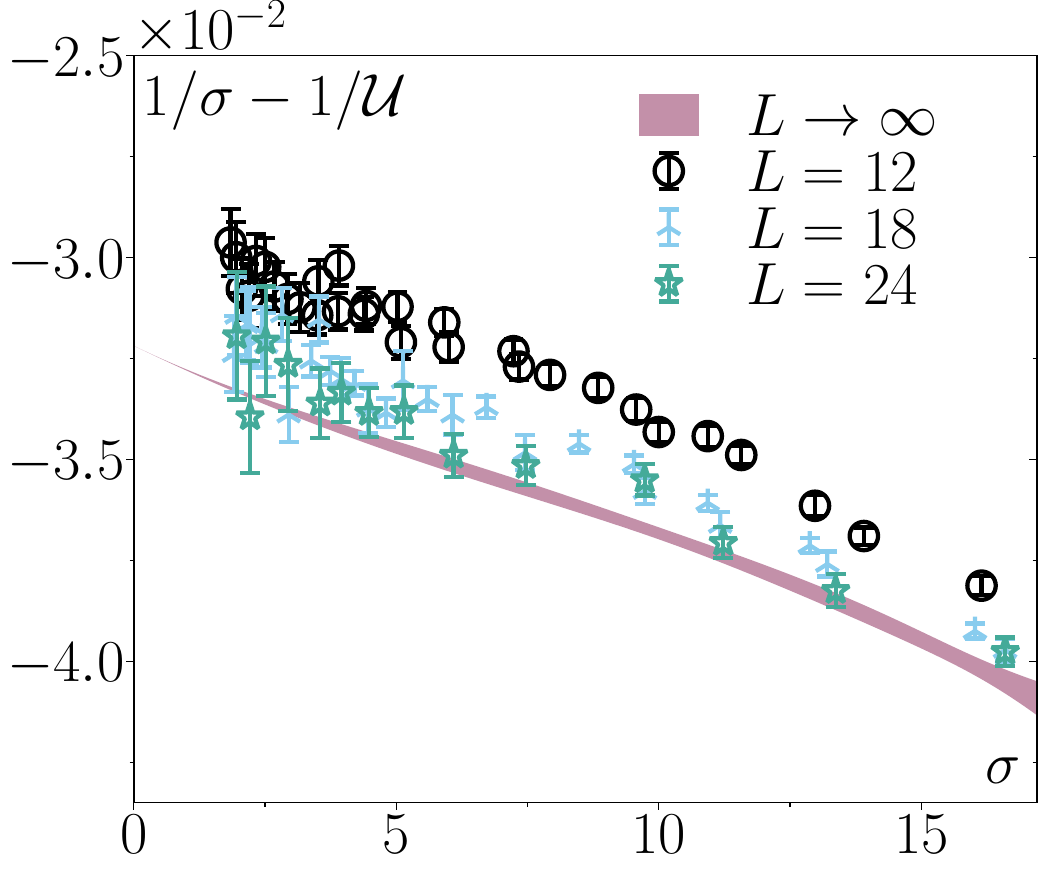}
\includegraphics[scale=0.33]{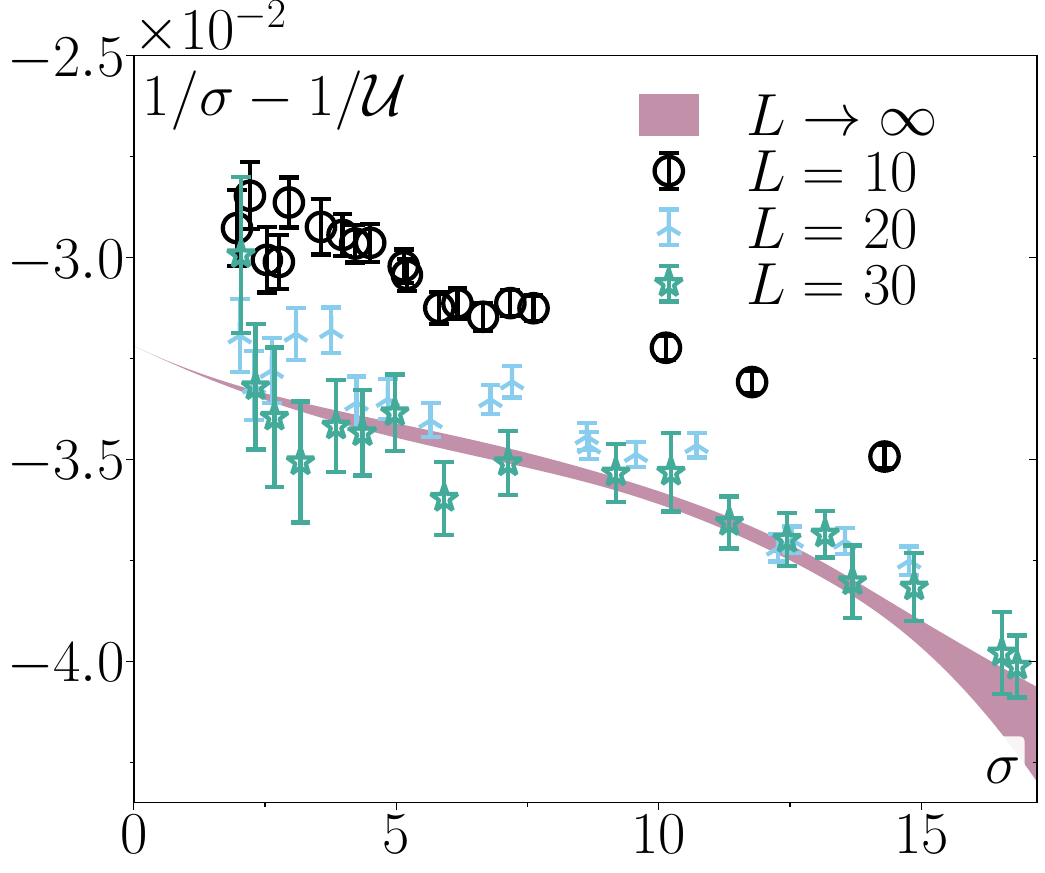}
\includegraphics[scale=0.33]{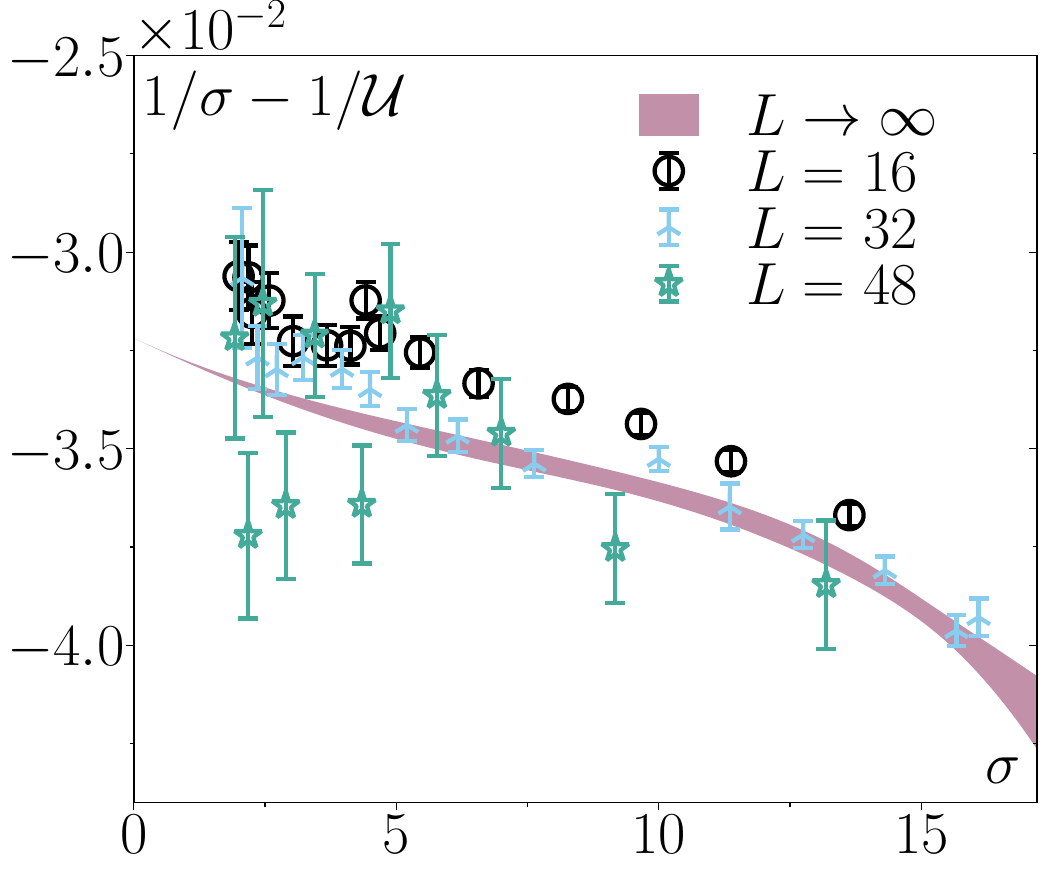}
\caption{Lattice (points) and continuum (shaded areas) step-scaling functions. From left to right: $N=3, 5, 8$.}
\label{fig:step_scaling_continuum}
\end{figure*}

If the renormalized coupling $\sigma$ is not too large, the continuum inverse step-scaling function $u(\sigma)$ can be expanded in powers of $\sigma$ up to some order $d_1$ as
\be\label{eq:inverse_step_function}
\begin{aligned}
\frac{1}{u(\sigma)} = \frac{1}{\sigma} + \sum_{n\,=\,0}^{d_1} p_n \sigma^n + \mathcal{O}(\sigma^{d_1+1})\,,
\end{aligned}
\ee
where the first two coefficients, $p_0 = 2b_0\log2$ and $p_1 = 2b_1\log2$, are constrained by perturbation theory. We assume that lattice artifacts in $\mathcal{U}(\sigma,L)$ are $(a\mu)^2 \propto 1 / L^2$ at leading order, and that the coefficient of these artifacts is a function of $\sigma$ that can be power-expanded up to some order $d_2$. This leads to
\be\label{eq:inverse_step_function_artifacts}
\frac{1}{{\cal U}(\sigma,L)}= \frac{1}{u(\sigma)} + \frac{1}{L^2} \sum_{n\,=\,0}^{d_2} \rho_n \sigma^n + \dots\,.
\ee
Combining \cref{eq:inverse_step_function} and \cref{eq:inverse_step_function_artifacts}, the parameterization of $u(\sigma)$ that we use in our global fits, one for each value of $N$, is
\be
\label{eq:step_global_fit}
\frac{1}{\mathcal{U}(\sigma,L)} - \frac{1}{\sigma} = \sum_{n\,=\,0}^{d_1} p_n \sigma^n + \frac{1}{L^2} \sum_{n\,=\,0}^{d_2}\rho_n \sigma^n \,.
\ee
If $d_1 = d_2=d$, \cref{eq:step_global_fit} could be written as
\be
\label{eq:step_global_fit_var}
\frac{1}{\mathcal{U}(\sigma,L)} - \frac{1}{\sigma} = \sum_{n\,=\,0}^{d}\left(p_n + \frac{\rho_n}{L^2}\right) \sigma^n \, ,
\ee
that is, it could be interpreted as assuming lattice artifacts of order $(a\mu)^2$ in the parameters of the expansion in \cref{eq:inverse_step_function}. However, we allow for $d_1 \neq d_2$ because, at a given order $n$ in $\sigma$, the parameter $p_n$ could be significantly larger or smaller than its lattice artifacts. In general, imposing fewer constraints on the fit parameters reduces the systematics of the chosen parameterization, at the cost of a larger uncertainty in the results of the fit. In the end, we quote results obtained with $d_1=d_2=4$ for $N=3,8$ and with $d_1=4, d_2=5$ for $N=5$ as our final determinations (best fit parameters parameterizing the continuum step scaling function can be found in Appendix~\ref{appendix:fits}). These values were chosen after checking the stability of the continuum result by separately varying $d_1$ and $d_2$ between 4 and 6. In all cases, we fit the couplings in the range $\lambda \in [1.5,18]$, corresponding for all $N$ to the range where the step-scaling function has a monotonic behavior that can be parameterized with a single Padé approximant.\footnote{
The \SU{3} simulation data are largely inherited from Ref.~\cite{Bribian:2021cmg}, but in the present paper we have reanalyzed them using the common prescription adopted for the analysis of the other values of $N$. In particular, the global fit in Ref.~\cite{Bribian:2021cmg} used the range \(\lambda\in[1.5,17]\) and starts the step-scaling sequence from $u_0=34.52$, whereas here we use \(\lambda\in[1.5,18]\) and $u_0 = 44.81$ for all three values of $N$. We also have increased statistics for a few simulation points, see~\cite{Bonanno:2024nba}. This explains the small numerical differences in the global-fit coefficients reported here and in~\cite{Bribian:2021cmg} for $N=3$.
}
The number of data points entering the fits and the corresponding reduced $\chi^2$ values for our final determinations of the step-scaling functions are reported in \cref{tab:params_global_fit}.

\begin{table}[!t]
\centering
\begin{tabular}{ccc}
\toprule
$N$ & $n_{\rm pt}$ & $\chi^2/{\rm dof}$ \\
\midrule
3 & 74 & 0.70 \\
5 & 55 & 1.16 \\
8 & 31 & 0.91 \\
\bottomrule
\end{tabular}
\caption{Number $n_{\rm pt}$ of data points included in the global fits of the step-scaling functions and corresponding values of the $\chi^2$ per degree of freedom (dof).}
\label{tab:params_global_fit}
\end{table}

The resulting continuum step-scaling functions are shown in \cref{fig:step_scaling_continuum}. The comparison between $N=3,5,8$ shows that the running of the 't Hooft coupling has a mild dependence on $N$, consistent with the expected large-$N$ scaling. The \SU{8} determination carries the largest uncertainty, especially in the weak-coupling region, because it is the most computationally demanding. 

From our previous studies~\cite{Bribian:2021cmg,Bonanno:2024nba}, we had at our disposal some simulation points of \SU{3} tuned to achieve LCPs in the renormalized coupling. Thus, we could also perform the continuum extrapolations of $\mathcal{U}(\sigma,L)$ at fixed value of $\sigma$, that is, without any assumption on the $\sigma$-dependence of $u(\sigma)$ and of its lattice artifacts. This approach and the global fit are found to give perfectly compatible results.

\subsection{Extraction of \texorpdfstring{$\Lambda/\muhad$}{Lambda/muhad}}
\label{subsec:LambdaTGF}

\begin{table*}[!t]
\centering
\begin{tabular}{ccccccc}
\toprule
& \multicolumn{2}{c}{\SU{3}} & \multicolumn{2}{c}{\SU{5}} & \multicolumn{2}{c}{\SU{8}} \\
\cmidrule(lr){2-3} \cmidrule(lr){4-5} \cmidrule(lr){6-7}
$k$ 
& $\lambda(2^{k}\muhad)$ & $\LambdaMS/(2\muhad)$
& $\lambda(2^{k}\muhad)$ & $\LambdaMS/(2\muhad)$
& $\lambda(2^{k}\muhad)$ & $\LambdaMS/(2\muhad)$\\
1	&	16.074(22)	&	0.22195(31)	&	16.014(39)      	&	0.21844(57)	&	16.04(21)	&	0.2183(28)	\\
2	&	9.769(20)	&	0.22125(86)	&	9.725(64)       	&	0.2168(27)	&	9.752(66)	&	0.2174(27)	\\
3	&	7.189(11)	&	0.22419(89)	&	7.212(32)       	&	0.2233(26)	&	7.218(37)	&	0.2232(30)	\\
4	&	5.724(11)	&	0.2268(14)	&	5.761(24)       	&	0.2290(32)	&	5.755(29)	&	0.2277(39)	\\
5	&	4.768(11)	&	0.2288(21)	&	4.805(20)       	&	0.2334(39)	&	4.795(25)	&	0.2309(49)	\\
6	&	4.093(10)	&	0.2305(28)	&	4.127(17)       	&	0.2370(47)	&	4.115(22)	&	0.2333(60)	\\
7	&	3.5887(95)	&	0.2318(35)	&	3.619(15)       	&	0.2398(54)	&	3.607(19)	&	0.2352(69)	\\
8	&	3.1976(86)	&	0.2329(40)	&	3.224(13)       	&	0.2422(61)	&	3.213(17)	&	0.2367(78)	\\
9	&	2.8848(78)	&	0.2338(45)	&	2.908(11)       	&	0.2442(67)	&	2.898(15)	&	0.2379(86)	\\
10	&	2.6288(70)	&	0.2346(49)	&	2.6488(100)     	&	0.2459(72)	&	2.640(13)	&	0.2390(92)	\\
11	&	2.4152(63)	&	0.2353(53)	&	2.4329(89)      	&	0.2473(76)	&	2.425(12)	&	0.2398(99)	\\
12	&	2.2343(57)	&	0.2359(56)	&	2.2500(79)      	&	0.2485(80)	&	2.243(11)	&	0.2405(104)	\\
13	&	2.0789(52)	&	0.2364(59)	&	2.0930(71)      	&	0.2496(84)	&	2.0862(95)	&	0.2412(109)	\\
14	&	1.9441(47)	&	0.2368(61)	&	1.9567(64)      	&	0.2506(87)	&	1.9505(86)	&	0.2417(114) \\	
15	&	1.8259(43)	&	0.2372(64)	&	1.8373(57)      	&	0.2515(90)	&	1.8316(78)	&	0.2422(118) \\	
$\infty$ & 0.0 & 0.2431(98) & 0.0 & 0.264(13) & 0.0 & 0.250(18) \\
\bottomrule
\end{tabular}
\caption{Step-scaling sequence of the coupling for \SU{3}, \SU{5} and \SU{8}, starting at $\lambda(2\muhad)$. The final result, at $k\to\infty$, is obtained from a linear extrapolation of the last five points in each sequence.}
\label{tab:lambda_muhad}
\end{table*}

\begin{figure}[t]
\centering
\includegraphics[scale=0.45]{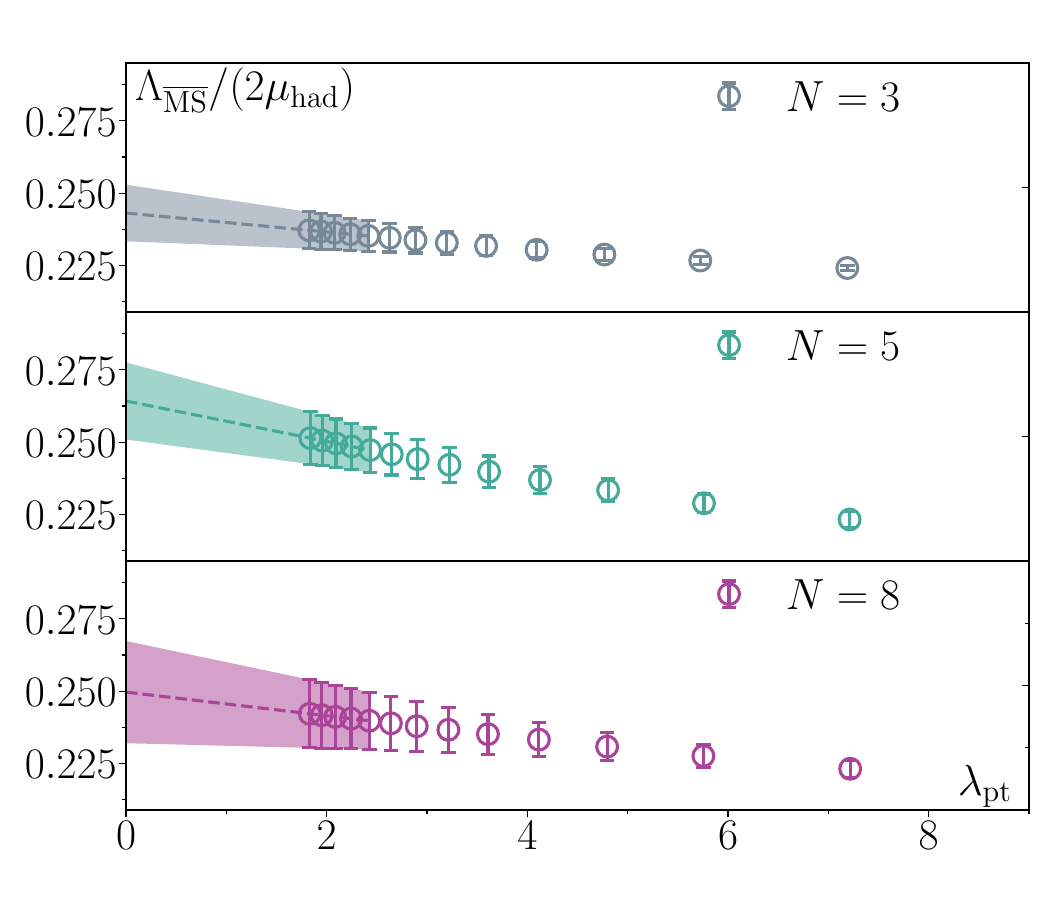}
\caption{Extrapolation towards $\lambda_{\pt} \to 0$ of $\LambdaMS/\muhad$ determined from step scaling as a function $\lambda_{\pt}$, the coupling at which the step-scaling procedure is matched to perturbation theory for $N=3,5,8$.}
\label{fig:lambda_extrapolation}
\end{figure}

Having determined the continuum step-scaling function $u(\sigma)$, we use it to build the sequence of couplings in \cref{eq:step-scaling-iteration} connecting the starting hadronic scale $\muhad$, implicitly defined by the chosen $u_0 = \lambda(\muhad)$, to the perturbative regime. The first step $u_1$ in the sequence has been determined in \cref{subsec:initial}. Starting from $u_1$, we generate the sequence $u_k = \lambda(2^k\muhad)$ by recursively applying $u(\sigma)$. At each step $k$ of the sequence, we use $u_k = \lambdaTGF(\mupt) \equiv \lambda_{\pt}$ with $\mupt = 2^k\muhad$ to compute
\be
\label{eq:lambda_sequence}
\frac{\LambdaTGF}{2\muhad}\bigg\vert_k = 2^{k-1} \frac{\LambdaTGF}{\mupt}\bigg\vert_{\mathrm{pt}} \,,
\ee
where $\frac{\LambdaTGF}{\mupt}\big\vert_{\mathrm{pt}}$ is evaluated with the perturbative expression of \cref{eq:Lambda_perturbative} using the two-loop universal $\beta$ function. At each step, the conversion to the $\MStext$ scheme is performed using the known exact one-loop matching factor:
\be
\label{eq:lambda_ms}
\frac{\LambdaMS}{2\muhad} = \frac{\LambdaMS}{\LambdaTGF} \times \frac{\LambdaTGF}{2\muhad}.
\ee

The final result is obtained via a linear extrapolation of $\LambdaMS/\muhad$ as a function of $\lambda_\pt$ towards $\lambda_\pt \to 0$, performed considering the last five points of the sequence, see Fig.~\ref{fig:lambda_extrapolation}. Since these values are statistically correlated, the linear extrapolation is done independently for each of the 20 jackknife samples we have generated. The central value is obtained from the jackknife average, while the statistical uncertainty is determined from the jackknife variance of the extrapolated intercepts.

\subsection{Determination of \texorpdfstring{$\LambdaMS\sqrt{8t_0}$}{LamdbaMSsqrt(8t0)}}\label{subsec:muhad_sqrt8t0}

As anticipated in \cref{subsec:step_scaling_lattice}, in order to obtain our final results for the $\Lambda$-parameter from \cref{eq:Lambda_with_step_scaling}, the hadronic scale $\muhad$ must be expressed in units of a conventional reference scale, which we chose to be the gradient flow scale $t_0$. Gradient-flow scales like $t_0$, $t_1$ and $w_0$ are defined through the behavior of the energy density $E(t)$ of the flowed gauge fields~\cite{Luscher:2010iy,Butti:2022sgy, BMW:2012hcm}. In particular, $t_0$ is defined in the continuum by the condition
\be
    \label{eq:t0}
    \frac{N}{N^2-1}\langle t^2 E(t) \rangle \bigg\vert_{t\,=\,t_0}= \frac{9}{80} = 0.1125 \, ,
\ee
which generalizes the standard \SU{3} definition~\cite{Luscher:2010iy} for arbitrary $N$, while ensuring a finite large-$N$ limit. In physical units, this corresponds to $\sqrt{8t_0}\simeq 0.475$ fm for \SU{3}~\cite{Giusti:2018cmp}, a conversion factor that we also adopt to express the lattice spacing in physical units for arbitrary $N$. On the lattice, $t_0$ is obtained from \cref{eq:t0} using the same clover energy discretization employed for the coupling.

Recalling that the renormalization scale in the TGF scheme is $\mu = (c\ell)^{-1}$ with $c=0.3$, the conversion factor in \cref{eq:muhadsqrt8t0_1}, needed to convert $\LambdaMS/\muhad$ to $\LambdaMS\sqrt{8t_0}$, is obtained from
\be\label{eq:muhadsqrt8t0}
a\muhad \times \sqrt{\frac{8t_0}{a^2}} = \frac{1}{cL} \times \sqrt{\frac{8t_0}{a^2}}\, .
\ee
This quantity must be calculated, for each value of $N$, for a set of lattice sizes $L$ and lattice spacings $a(b)$ such that $\lambda(b, L) = u_0$, and then extrapolated to the continuum. The sets of bare couplings corresponding to the same $u_0$ can be extracted from the Padé fits described in \cref{subsec:initial}.

Unlike step scaling, scale setting requires additional simulations in large volumes to have controlled infinite-volume extrapolations. These were done in Ref.~\cite{Bonanno:2025kfd}, where we determined $t_0$ and other gradient-flow scales for the sets of values of $b$ giving the LCP with renormalized coupling $u_0$ for each value of $N$. This was done by combining \TBC and the \emph{Parallel Tempering on Boundary Conditions} (PTBC) algorithm~\cite{Hasenbusch:2017unr,Bonanno:2020hht,Bonanno:2024zyn}, which allows to significantly reduce the autocorrelation time of the topological charge compared to standard algorithms based on periodic or open boundary conditions, see~\cite{Bonanno:2025eeb}. This was necessary to tame the severe topological freezing affecting standard simulations at the fine lattice spacings we are considering. We used several volumes to achieve a controlled infinite-volume extrapolation, both for the standard and the tree-level improved definitions of the energy density entering the definition of $t_0$. As a cross-check, we verified that our results for $t_0$ for $N=3$ were perfectly compatible with previous ones obtained for the same bare couplings via Master Field simulations~\cite{Giusti:2018cmp}. The corresponding lattice spacings range for all $N$ from approximately $a \sim 0.065\,\mathrm{fm}$ down to $a \sim 0.025\,\mathrm{fm}$, a range that enables a controlled continuum limit of $\muhad$. The relevant scale-setting results are reported in \cref{tab:scale_setting}. For $N=5$ and $8$, we use our determinations from Ref.~\cite{Bonanno:2025kfd}, while for $N=3$ we use an interpolation of those from Ref.~\cite{Giusti:2018cmp}.

\begin{figure}[!t]
\centering
\includegraphics[scale=0.45]{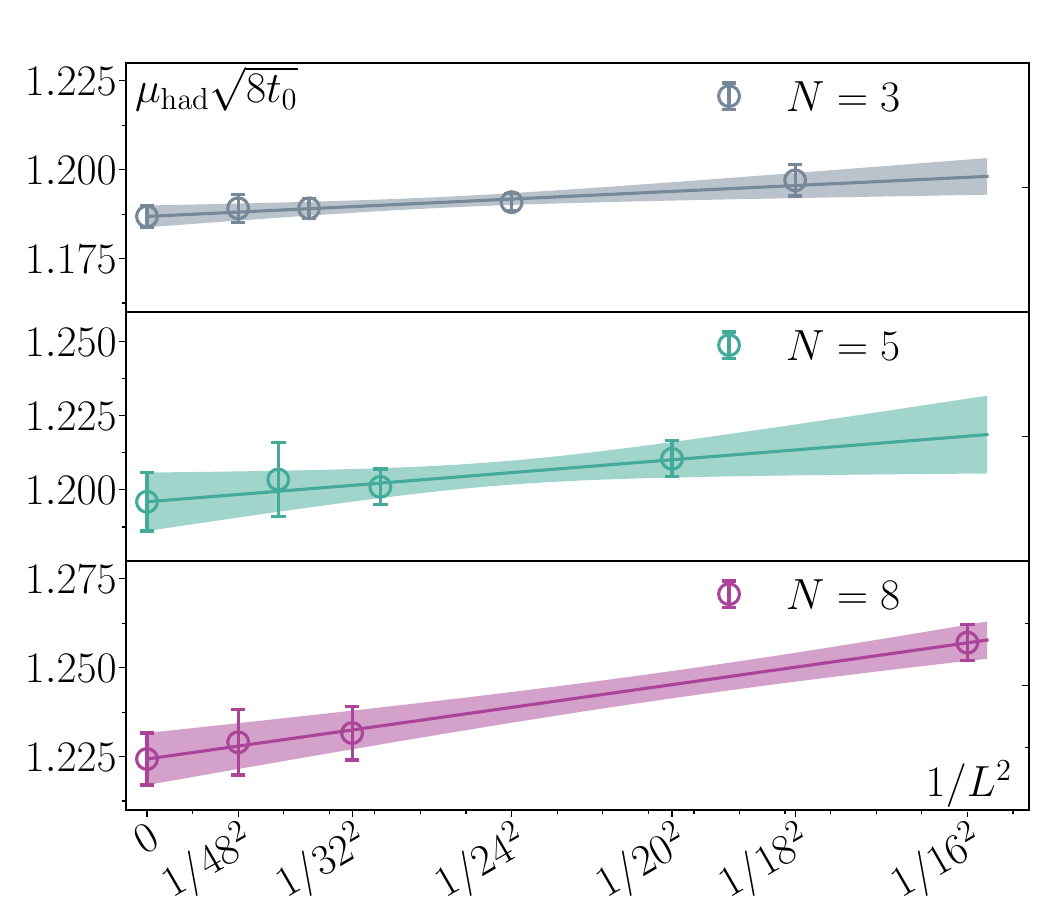}
\caption{Determination of the conversion factor $\muhad\sqrt{8t_0}$ for $N=3, 5, 8$, from top to bottom. The continuum limit is determined assuming $\mathcal{O}(a^2)$ lattice artifacts, proportional to $1/L^2$ at fixed $\muhad$. The results are reported in \cref{tab:final_res}.}
\label{fig:muhad_sqrt8t0}
\end{figure}

The continuum limit of the conversion factor in \cref{eq:muhadsqrt8t0} is determined assuming $\mathcal{O}[(a\muhad)^2]=\mathcal{O}(1/L^2)$ lattice artifacts, as shown in \cref{fig:muhad_sqrt8t0}. The error associated to the tuning of the bare couplings to match the LCP determined by $u_0$, is propagated to $\muhad$ via a bootstrap analysis and a smooth interpolation of $\log\{t_0/a^2\}$ as a function of $b$. The results are reported in \cref{tab:final_res}, which also contains the final results for $\LambdaMS\sqrt{8t_0}$ for $N$=3, 5 and 8, obtained as
\be
    \label{eq:Lambda_phys_with_step_scaling}
    \LambdaMS\sqrt{8t_0} = 2\times\frac{\LambdaMS}{2\muhad}\times\muhad \sqrt{8t_0}\, .
\ee
The result obtained for $N=3$, $\LambdaMS\sqrt{8t_0}= 0.577 (23)$, agrees with the one determined in Ref.~\cite{Bribian:2021cmg}. Our analysis differs slightly from the one presented there, since we have changed the  choice of the hadronic reference scale from $\muhad \sqrt{8t_0} =1.3860(62)$ to $1.1868(30)$, in order to match the procedure followed for $N=5$, and 8, as described in \cref{subsec:initial}. Moreover, in Ref.~\cite{Bribian:2021cmg}, the TGF coupling was matched non-perturbatively to the Schr\"odinger Functional scheme, for which the 3-loop $\beta$ function is known. This allows a better control of the perturbative extrapolation in \cref{fig:lambda_extrapolation} and leads to a result: $\LambdaMS\sqrt{8t_0} = 0.603(17)$, perfectly compatible within errors with the one based only on the TGF scheme.

\begin{table}[!t]
\centering
\begin{tabular}{ccccc}
\midrule
$N$ & $L$& $b$ & $\beta_g$ & $t_0/a^2$ \\
\midrule
\multirow{4}{*}[-0.0em]{3} & 18 &  0.34184  &  6.153 &  5.2227(84)  \\
                           & 24 &  0.35267  &  6.348 &  9.190(15)   \\
                           & 36 &  0.36926  &  6.647 &  20.617(33)  \\
                           & 48 &  0.38166  &  6.870 &  36.652(59)  \\
\addlinespace[0.5em]
\multirow{3}{*}[-0.0em]{5} & 20 &  0.35971  & 17.985 &   6.595(12)  \\
                           & 30 &  0.37504  & 18.752 &  14.606(40)  \\
                           & 40 &  0.38683  & 19.342 &  26.067(99)  \\
\addlinespace[0.5em]
\multirow{3}{*}[-0.0em]{8} & 16 &  0.35867  & 45.910 &  4.5507(74)  \\
                           & 32 &  0.38352  & 49.091 &  17.470(34)  \\
                           & 48 &  0.40008  & 51.210 &   39.16(16)  \\
\bottomrule
\end{tabular}
\caption{Scale setting determinations needed to compute $\muhad\sqrt{8t_0}$. Data for $N=3$ come from a spline interpolation of the results of Ref.~\cite{Giusti:2018cmp}. Data for $N=5,8$ come instead from Ref.~\cite{Bonanno:2025kfd}.}
\label{tab:scale_setting}
\end{table}

\begin{figure}[!t]
\centering
\includegraphics[scale=0.45]{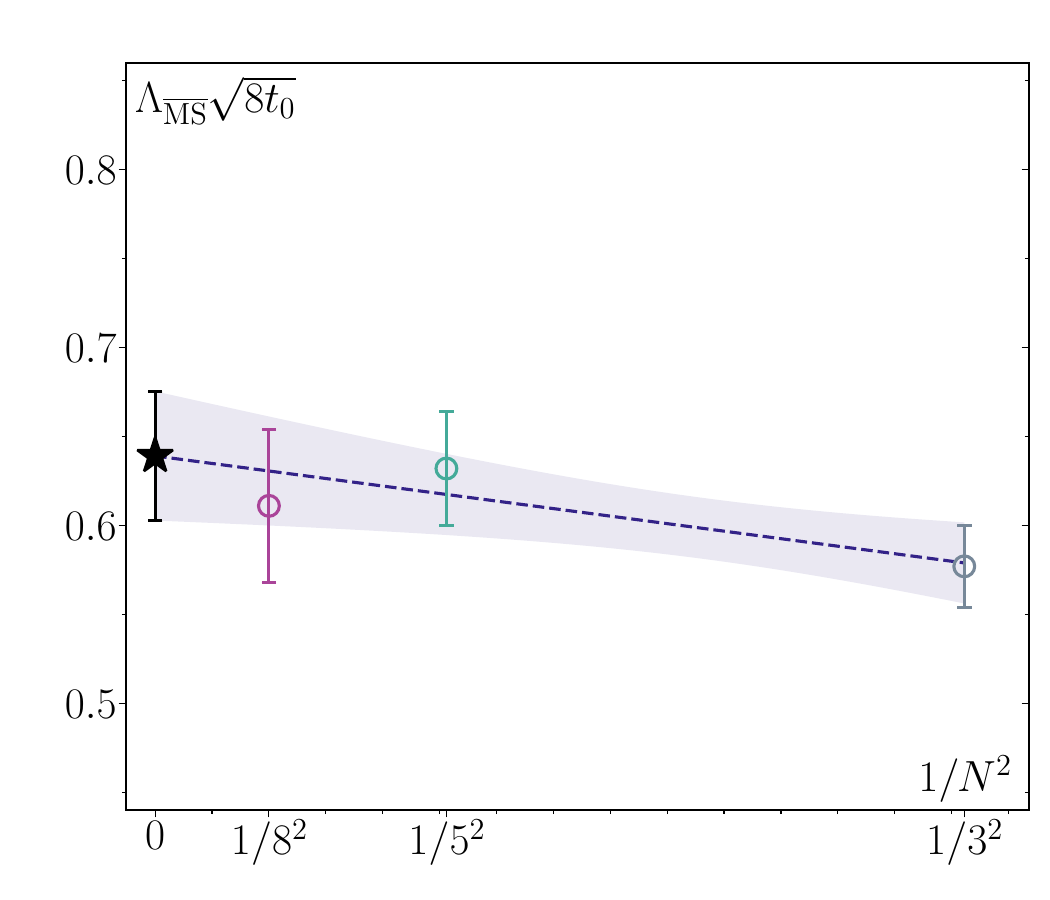}
\caption{Linear fit of the $N$-dependence of our determinations of $\LambdaMS \sqrt{8t_0}$. The large-$N$ extrapolation is $\LambdaMS\sqrt{8t_0} = 0.639(36)$ and the slope coefficient in \cref{eq:Lambda_largeN_fit} is $A = -0.85(62)$.}
\label{fig:largeN_Lambda}
\end{figure}

\begin{table}[!t]
\centering
\begin{tabular}{cccc}
\midrule
$N$ & $\LambdaMS/(2\muhad)$ & $\muhad\sqrt{8t_0}$ & $\LambdaMS \sqrt{8t_0}$ \\
\midrule
3        & 0.2431(98) & 1.1868(30)  & 0.577(23) \\
5        & 0.264(13)  & 1.1959(99)  & 0.632(32) \\
8        & 0.250(18)  & 1.2243(73)  & 0.611(43) \\
$\infty$ & --         & --          & 0.639(36) \\
\bottomrule
\end{tabular}
\caption{Final results for the $\Lambda$ parameter in units of $\muhad$, along with the corresponding determinations of the hadronic scale $\muhad$ in units of $\sqrt{8t_0}$.}
\label{tab:final_res}
\end{table}

\subsection{The large-\texorpdfstring{$N$}{N} limit of \texorpdfstring{$\sqrt{8t_0}\LambdaMS(N)$}{sqrt(8t0)LambdaMSbar(N)}}
\label{subsec:Lambda_physical}

This section is devoted to the calculation of the $N$-dependence and of the large-$N$ limit of our determinations of the $\Lambda$-parameter. These are obtained from a large-$N$ extrapolation of $\sqrt{8t_0}\LambdaMS(N)$, obtained via a best fit of our $N=3,5,8$ determinations assuming the following $1/N$ expansion:
\be\label{eq:Lambda_largeN_fit}
\sqrt{8t_0}\LambdaMS(N) = \sqrt{8t_0}\LambdaMS(\infty)
\left[1 + \frac{A}{N^2} +\mathcal{O}\left(\frac{1}{N^4}\right)\right] \, .
\ee
The results of the best fit, shown in \cref{fig:largeN_Lambda}, are:
\be\label{eq:largeN_lambda_results}
\sqrt{8t_0}\LambdaMS(\infty) = 0.639(36) \,, \quad A = -0.85(62) \,.
\ee
The large-$N$ limit turns out to be pretty smooth, with the $N=\infty$ determination being just $\sim 11\%$ larger than the $N=3$ result. Our final uncertainties on $\sqrt{8t_0}\LambdaMS(N)$ are about $4.0\%,\, 5.1\%,\, 7.0\%, 5.6\%$ for $N=3,\,5,\,8,\infty$.

\section{Comparison with the literature and conclusions}\label{sec:conclusions}

\begin{figure}[!t]
\centering
\includegraphics[width=\columnwidth]{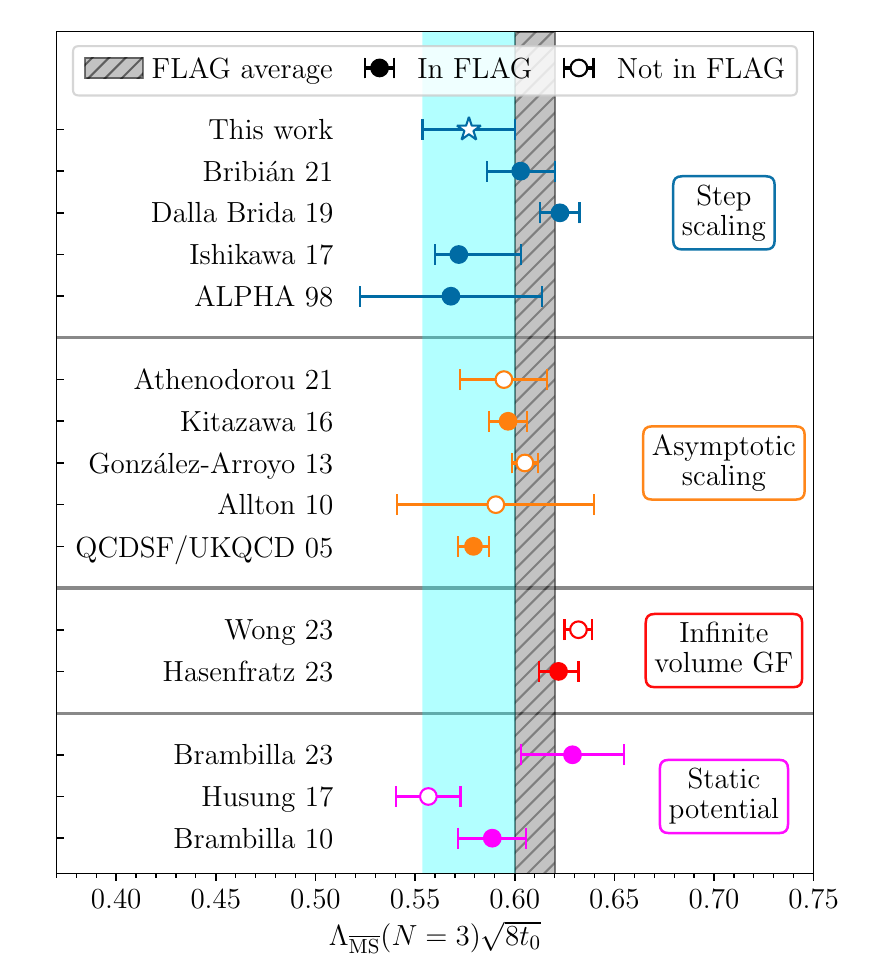}
\caption{Comparison of our $\LambdaMS$ result for \SU{3} with literature. The results entering the FLAG average~\cite{FlavourLatticeAveragingGroupFLAG:2024oxs} (grey vertical band) are shown as filled points and with their original errors, not the actual weights assigned by FLAG. Results are grouped by the method used for the determination: step scaling for Refs.~\cite{Bribian:2021cmg,DallaBrida:2019wur,Ishikawa:2017xam,CAPITANI1999669}, asymptotic scaling for Refs.~\cite{Athenodorou:2021qvs,Kitazawa:2016dsl,Gonzalez-Arroyo:2012euf,Allton:2008ty,PhysRevD.73.014513}, infinite-volume gradient flow scheme for Refs.~\cite{Wong:2023CY,Hasenfratz:2023bok}, and static-potential for Refs.~\cite{PhysRevD.109.114517,Husung:2017qjz,Brambilla:2010pp}.}
\label{fig:Lambda_FLAG_comp_N3}
\end{figure}

\begin{figure}[!t]
\centering
\includegraphics[width=\columnwidth]{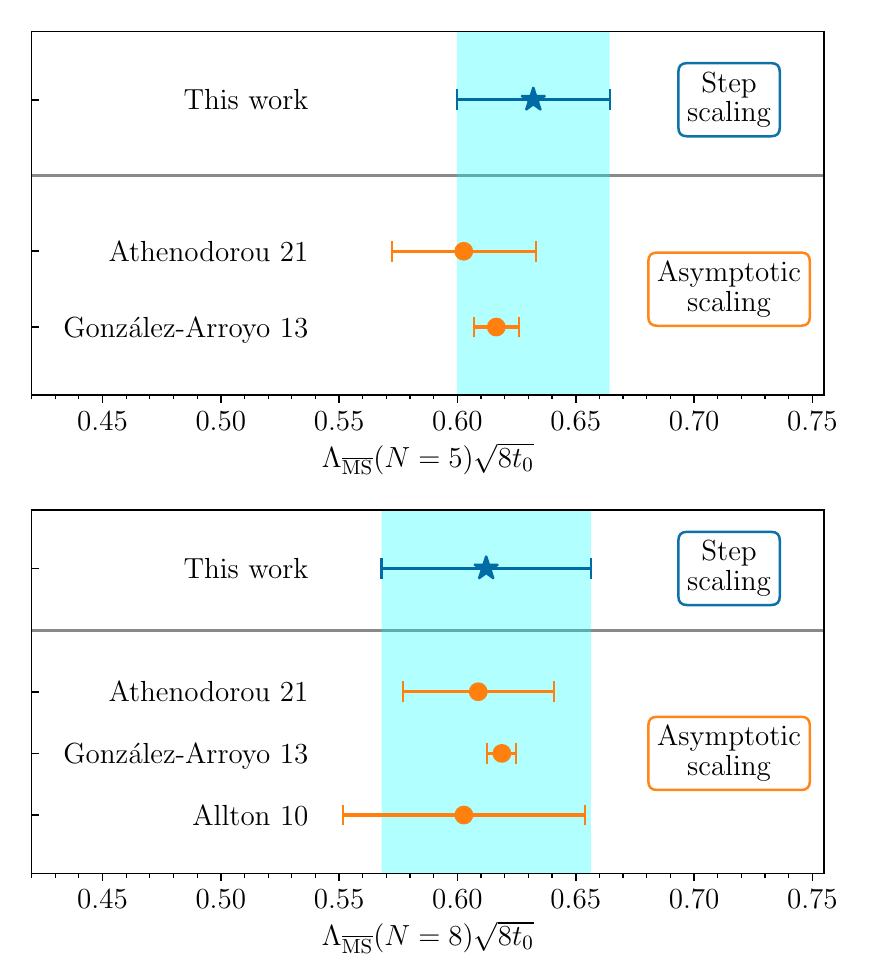}
\caption{Comparison of our $\LambdaMS$ results for \SU{5} (top panel) and \SU{8} (bottom panel) with those from Refs.~\cite{Athenodorou:2021qvs,Gonzalez-Arroyo:2012euf} (for \SU{5} and \SU{8}) and Ref.~\cite{Allton:2008ty} (for \SU{8}).}
\label{fig:Lambda_FLAG_comp_N5_N8}
\end{figure}

\begin{figure}[!t]
\centering
\includegraphics[width=\columnwidth]{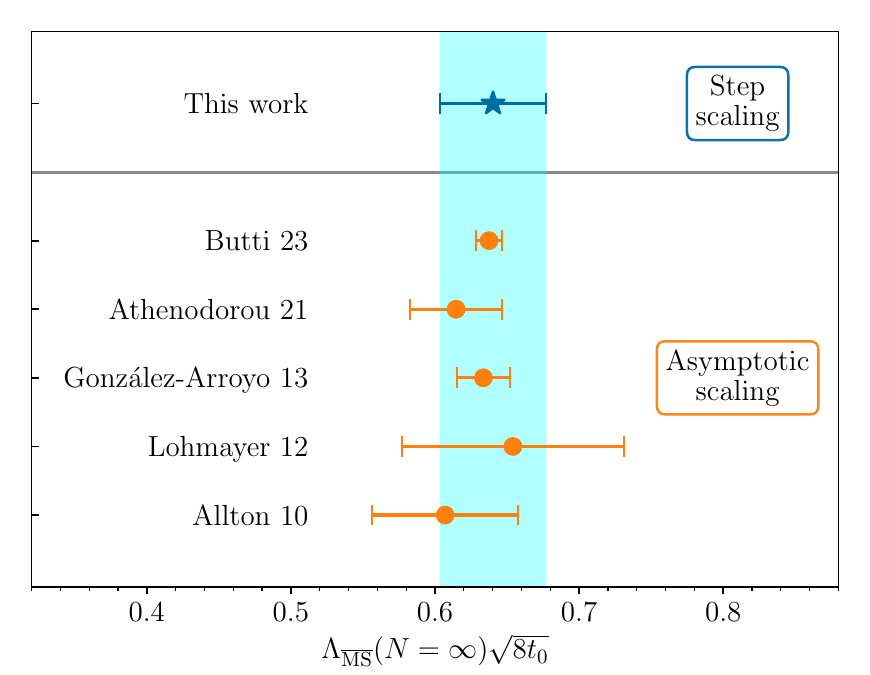}
\caption{Comparison of our large-$N$ limit of $\LambdaMS$ with those of Refs.~\cite{Butti:2023hfp,Athenodorou:2021qvs,Gonzalez-Arroyo:2012euf,Lohmayer:2012ue,Allton:2008ty}.}
\label{fig:Lambda_largeN_comp_Ninf}
\end{figure}

Let us now conclude our study by comparing our results with other determinations available in the literature, all collected in \cref{tab:Lambda-literature_COMP}. For the comparison, we express $\LambdaMS$ in units of $t_0$ in all cases because this quantity is determined more accurately than the string tension $\sigma$ or the Sommer parameter $r_0$. We also classify the results by the strategy employed to obtain them.

The comparison for \SU{3} is shown in \cref{fig:Lambda_FLAG_comp_N3}, where we also show the average $\LambdaMS\sqrt{8t_0} = 0.610(10)$ reported by FLAG~\cite{FlavourLatticeAveragingGroupFLAG:2024oxs}. Results from literature are shown with the uncertainty originally assigned by the respective authors, not those used by FLAG in the average. The latter are from $5\%$ to $7\%$ larger than the former for the determinations using the infinite-volume gradient-flow scheme (Refs.~\cite{Wong:2023CY,Hasenfratz:2023bok}) and asymptotic scaling (Refs.~\cite{Kitazawa:2016dsl,PhysRevD.73.014513}). Also, we only show the quadrature-sum of the statistical and systematic errors. In particular, the uncertainty on Kitazawa 16~\cite{Kitazawa:2016dsl} includes the 1\% systematic error that the authors estimate as the possible bias of topological freezing on their determination. Results originally given in units of the string tension $\sigma$ are converted to $t_0$ using $\sqrt{8t_0\sigma} = 1.0962(46)$ for $N=3$ from Ref.~\cite{Bonanno:2025eeb}, obtained from a continuum extrapolation of data for $t_0$~\cite{Ce:2015qha} and $\sigma$~\cite{Athenodorou:2020ani}. Results originally given in units of the Sommer parameter $r_0$ are converted to $t_0$ using $\sqrt{8t_0}/r_0 = 0.9435(97)$ from Ref.~\cite{DallaBrida:2019wur}. The results of Refs.~\cite{Gonzalez-Arroyo:2012euf,Kitazawa:2016dsl}, originally given in units of the gradient-flow scales $t_1,\,w_0$ respectively, are converted using $t_1/t_0 = 0.4204(11)$ and $w_0^2/t_0 = 1.0425(54)$ from our scale-setting study in Ref.~\cite{Bonanno:2025kfd}. In Ishikawa 17~\cite{Ishikawa:2017xam}, $\LambdaMS$ is determined both in units of $\sigma$ and $r_0$. We get $\LambdaMS\sqrt{8t_0} = 0.583(9)[^{+29}_{-5}]$ converting their result given in units of $\sigma$, and $\LambdaMS\sqrt{8t_0} = 0.572(9)[^{+29}_{-5}]$ converting from $r_0$.

As discussed in \cref{subsec:muhad_sqrt8t0}, our result for $N=3$ cannot be considered an independent determination from Bribián 21~\cite{Bribian:2021cmg}. The latter is based on the same set of numerical simulations but goes through a non-perturbative matching between the TGF and SF couplings to obtain a better control of the perturbative extrapolation.
Our current $\sim 4\%$ error is about twice the uncertainty on the world average, and our result is both compatible with the most precise step-scaling and asymptotic-scaling determinations by Dalla Brida 19~\cite{DallaBrida:2019wur} and QCDSF/UKQCD 05~\cite{PhysRevD.73.014513}. Thus, we are not yet able to address the existing tensions. Since a significant reduction of the uncertainties would be necessary to this end, this is left for a future dedicated work.

\begin{table}[!t]
\footnotesize
\begin{tabular}{@{\hskip -0.2cm} ccccc @{\hskip -0.2cm}}
\toprule
\multicolumn{5}{c}{$N=3$} \\
\midrule
$s$ & $\sqrt{8t_0}/s$ & Work & $\LambdaMS s$ & $\LambdaMS\sqrt{8t_0}$ \\
\midrule
\multirow{6}{*}{$\sqrt{8t_0}$} & \multirow{6}{*}{1} &  This work                                 &  0.577(23)   &  0.577(23)  \\
                               &                    &  Brambilla 23~\cite{PhysRevD.109.114517}   &  0.629(26)   &  0.629(26)  \\
                               &                    &  Hasenfratz 23~\cite{Hasenfratz:2023bok}   &  0.622(10)   &  0.622(10)  \\
                               &                    &  Wong 23~\cite{Wong:2023CY}                &  0.6320(70)  &  0.6320(70) \\
                               &                    &  Bribi\'an 21~\cite{Bribian:2021cmg}       &  0.603(17)   &  0.603(17)  \\
                               &                    &  Dalla Brida 19~\cite{DallaBrida:2019wur}  &  0.6227(98)  &  0.6227(98) \\
\addlinespace[0.5em]
\multirow{4}{*}{$\frac{1}{\sqrt{\sigma}}$} & \multirow{4}{*}{\makecell{1.0962(46)\\\cite{Bonanno:2025eeb}}} &  Athenodorou 21~\cite{Athenodorou:2021qvs}            &   0.542(20)  &     0.595(22) \\
                                           &                                                                &  Ishikawa 17*~\cite{Ishikawa:2017xam}                  &  $0.532(8)[^{\scriptscriptstyle{+27}}_{\scriptscriptstyle{-5}}]$  & $0.583(9)[^{\scriptscriptstyle{+29}}_{\scriptscriptstyle{-5}}]$ \\
                                           &                                                                &  Gonz\'alez-Arroyo 13~\cite{Gonzalez-Arroyo:2012euf}  &  0.5520(56)  &  0.6051(66) \\
                                           &                                                                &  Allton 10~\cite{Allton:2008ty}                       &   0.539(45)  &  0.591(49) \\
\addlinespace[0.5em]
\multirow{2}{*}{$w_0$} & 2.7701(72)             & \multirow{2}{*}{Kitazawa 16~\cite{Kitazawa:2016dsl}} & \multirow{2}{*}{0.2154(34)} & \multirow{2}{*}{0.5967(94)} \\
                       & \cite{Bonanno:2025kfd} &                                                      &                             &                             \\
\addlinespace[0.5em]
\multirow{5}{*}{$r_0$} & \multirow{5}{*}{\makecell{0.9435(97)\\\cite{DallaBrida:2019wur}}} &  Ishikawa 17~\cite{Ishikawa:2017xam}       &  $0.606(9)[^{\scriptscriptstyle{+31}}_{\scriptscriptstyle{-5}}]$  & $0.572(10)[^{\scriptscriptstyle{+30}}_{\scriptscriptstyle{-7}}]$ \\
                       &                    &  Husung 17~\cite{Husung:2017qjz}                      &  0.590(16)   &  0.557(16)  \\
                       &                                                                   &  Brambilla 10~\cite{Brambilla:2010pp}                 &  0.624(17)   &  0.589(17)  \\
                       &                                                                   &  \makecell{QCDSF/UKQCD 05~\cite{PhysRevD.73.014513}}  &  0.6140(54)  &  0.5793(78) \\
                       &                                                                   &  ALPHA 98~\cite{CAPITANI1999669}                      &  0.602(48)   &  0.568(46)  \\
\bottomrule
\end{tabular}
\vspace{1.5em}
\begin{center}
\begin{tabular}{ccccc}
\toprule
\multicolumn{5}{c}{$N=5$} \\
\midrule
$s$ & $\sqrt{8t_0}/s$ & Work & $\LambdaMS s$ & $\LambdaMS\sqrt{8t_0}$ \\
\midrule
\multirow{1}{*}{$\sqrt{8t_0}$} & \multirow{1}{*}{1} &                                  This work  &    0.632(32)  &    0.632(32) \\
\addlinespace[0.5em]
\multirow{2}{*}{$\frac{1}{\sqrt{\sigma}}$} & \multirow{2}{*}{\makecell{1.1648(48)\\\cite{Bonanno:2025eeb}}} &              Athenodorou 21~\cite{Athenodorou:2021qvs}  &   0.517(26)  &     0.603(30) \\
                                                   &                                     &    Gonz\'alez-Arroyo 13~\cite{Gonzalez-Arroyo:2012euf}  &  0.5292(78)  &    0.6164(94) \\
\bottomrule
\end{tabular}
\end{center}
\vspace{0.5em}
\begin{center}
\begin{tabular}{ccccc}
\toprule
\multicolumn{5}{c}{$N=8$} \\
\midrule
$s$ & $\sqrt{8t_0}/s$ & Work & $\LambdaMS s$ & $\LambdaMS\sqrt{8t_0}$ \\
\midrule
\multirow{1}{*}{$\sqrt{8t_0}$} & \multirow{1}{*}{1} &                                  This work  &    0.611(43)  &     0.611(43) \\
\addlinespace[0.5em]
\multirow{3}{*}{$\frac{1}{\sqrt{\sigma}}$} & \multirow{3}{*}{\makecell{1.1903(37)\\\cite{Bonanno:2025eeb}}} &              Athenodorou 21~\cite{Athenodorou:2021qvs}  &   0.511(27)  &     0.609(32) \\
                                         &                                     &    Gonz\'alez-Arroyo 13~\cite{Gonzalez-Arroyo:2012euf}  &  0.5199(49)  &    0.6188(61) \\
                                         &                                     &                         Allton 10~\cite{Allton:2008ty}  &   0.506(43)  &     0.603(51) \\
\bottomrule
\end{tabular}
\end{center}
\vspace{0.5em}
\begin{center}
\begin{tabular}{ccccc}
\toprule
\multicolumn{5}{c}{$N=\infty$} \\
\midrule
$s$ & $\sqrt{8t_0}/s$ & Work & $\LambdaMS s$ & $\LambdaMS\sqrt{8t_0}$ \\
\midrule
\multirow{1}{*}{$\sqrt{8t_0}$} & \multirow{1}{*}{1} &                                  This work  &   0.639(36)  &     0.639(36) \\
\addlinespace[0.5em]
\multirow{2}{*}{$\sqrt{8t_1^{\prime}}$} & 1.7466(33)             & \multirow{2}{*}{Butti 23~\cite{Butti:2023hfp}} & \multirow{2}{*}{0.3650(50)} & \multirow{2}{*}{0.6375(88)} \\
                                        & \cite{Bonanno:2025kfd} &                                                &                             &                             \\
\addlinespace[0.5em]
\multirow{4}{*}{$\frac{1}{\sqrt{\sigma}}$} & \multirow{4}{*}{\makecell{1.2068(46)\\\cite{Bonanno:2025eeb}}} &              Athenodorou 21~\cite{Athenodorou:2021qvs}  &   0.509(26)  &     0.615(32) \\
                                         &                                     &    Gonz\'alez-Arroyo 13~\cite{Gonzalez-Arroyo:2012euf}  &   0.525(15)  &     0.634(18) \\
                                         &                                     &                     Lohmayer 12~\cite{Lohmayer:2012ue}  &   0.542(64)  &     0.654(77) \\
                                         &                                     &                         Allton 10~\cite{Allton:2008ty}  &   0.503(42)  &     0.607(51) \\
\bottomrule
\end{tabular}
\end{center}
\caption{Results for the \SU{N} $\Lambda$-parameter for $N=3,5,8,\infty$ from the literature, with conversion to the gradient-flow scale $t_0$. Results are grouped according to the reference scale $s$ used in the original determination. We only report the quadrature-sum of the statistical and systematic errors, except for the two results from Ishikawa 17 due to the largely asymmetric systematics. The starred \SU{3} result in units of $\sigma$ from Ishikawa 17 is included for completeness but omitted from \cref{fig:Lambda_FLAG_comp_N3}, where we only show the result obtained from the determination in units of $r_0$ from the same work.}
\label{tab:Lambda-literature_COMP}
\end{table}

Let us now move to the case $N>3$. A central aspect of this work is that it provides the first determination of the large-$N$ Yang--Mills $\Lambda$-parameter based on a fully non-perturbative step-scaling strategy, avoiding the use of asymptotic-scaling methods. The only other step-scaling result in the literature is the \SU{4} determination of~\cite{Lucini:2008vi}. In that study, the authors employed the SF scheme to determine the \SU{4} $\Lambda$-parameter, obtaining a result $\LambdaMS/\sqrt{\sigma}= 0.527(21)_{\scriptscriptstyle{\rm stat}}(62)_{\scriptscriptstyle{\rm syst}}$~\cite{Lucini:2008vi}. Converted to $t_0$ units using $\sqrt{8t_0\sigma}(N=4) = 1.1446(47)$~\cite{Bonanno:2025eeb} (again, obtained using data of~\cite{Ce:2015qha,Athenodorou:2020ani}), this leads to $\sqrt{8t_0}\LambdaMS = 0.603(24)_{\scriptscriptstyle{\rm stat}}(71)_{\scriptscriptstyle{\rm syst}}$, in good agreement with the result obtained from our fit to the large-$N$ dependence in \cref{eq:Lambda_largeN_fit}, which gives a value of $\sqrt{8t_0}\LambdaMS = 0.605(18)$ for $N=4$. Let us now turn to the comparison with asymptotic scaling results. In \cref{fig:Lambda_FLAG_comp_N5_N8} we show the comparisons for \SU{5} and \SU{8}. Results from other works, originally given in units of $\sigma$, are converted to $t_0$ using $\sqrt{8t_0\sigma} = 1.1648(48),\, 1.1903(37)$ for $N=5,\,8$ from Ref.~\cite{Bonanno:2025eeb}. The conversion factor for $N=5$ is the result of a continuum extrapolation of data for $t_0$~\cite{Ce:2016awn} and $\sigma$~\cite{Athenodorou:2021qvs}, while the $N=8$ value is obtained from the large-$N$ extrapolation of $\sqrt{8t_0\sigma}$ in Ref.~\cite{Bonanno:2025eeb}. For \SU{5} and \SU{8} we observe good agreement, albeit within our larger error bars, with previous asymptotic scaling predictions.

Finally, we show the comparison for the available large-$N$ estimates in \cref{fig:Lambda_largeN_comp_Ninf}. For Butti 23, we report an updated value, obtained through private communication~\cite{Butti:Lambda_forthcoming}, of the result in Ref.~\cite{Butti:2023hfp}. The result, originally given in units of the gradient-flow scale $t_1$ (which we call $t_1^\prime$), is converted using $t_1^\prime/t_0 = 0.3278(12)$ from our scale-setting study~\cite{Bonanno:2025kfd}. Other results, originally given in units of $\sigma$, are converted to $t_0$ using the large-$N$ extrapolated value $\sqrt{8t_0\sigma}\big\vert_{N\to\infty} = 1.2068(46)$ obtained in~\cite{Bonanno:2025eeb}. Within the attained accuracy, we again observe good agreement with asymptotic scaling predictions. The behavior observed for $N>3$ is thus in line with the one found for $N=3$, where no tension between the two methods is observed at the level of $\sim 5\%$.

Due to its phenomenological relevance, in the next future it would be interesting to perform a dedicated calculation of $\LambdaMS(N=3)$ adopting the framework presented in this study, and pushing the accuracy to the per-cent level. This cannot be easily achieved by just increasing statistics or pushing the simulations closer to the continuum limit, as the dominant contribution to the error budget comes from the $\lambda_\pt \to 0$ extrapolation, which causes an increase of about 50\% of the error on the determination of $\LambdaMS$ at our most perturbative couplings, see \cref{tab:lambda_muhad}. Taming this source of uncertainty requires either matching to an intermediate scheme, such as the Schr\"odinger functional one, where the perturbative expansion is better behaved~\cite{DallaBrida:2019wur}, or numerically computing the first non-universal coefficient of the $\beta$-function in the TGF scheme via NSPT techniques~\cite{DiRenzo:1994av,DiRenzo:1994sy,DiRenzo:2004hhl,DallaBrida:2017tru}, which would make the leading corrections to the $\lambda_\pt \to 0$ limit quadratic rather than linear in $\lambda_\pt$.

Another interesting future development would be to exploit the large-volume gradient-flow data generated in~\cite{Bonanno:2025kfd} for the determination of $\muhad \sqrt{8t_0}$ to compute the $N$-dependence of the $\Lambda$-parameter using techniques akin to those adopted in~\cite{Hasenfratz:2023bok,Wong:2023CY}, based on the infinite-volume gradient flow scheme.

Finally, an intriguing future outlook would be to provide a step-scaling determination of the $\Lambda$-parameter directly in the large-$N$ limit by discretizing the theory on a fully-reduced 1-point twisted box. This setup, the so-called twisted Eguchi--Kawai formulation, has been adopted in a series of recent lattice studies of large-$N$ gauge theories, see Refs.~\cite{Gonzalez-Arroyo:2012euf,Gonzalez-Arroyo:2013bta,GarciaPerez:2014azn,GarciaPerez:2015rda,Gonzalez-Arroyo:2015bya,Perez:2020vbn,Butti:2022sgy,Butti:2023hfp,Bonanno:2023ypf,Bonanno:2024bqg,Bonanno:2024onr,Bonanno:2025hzr,Bonanno:2025bla,Hamada:2025whg}. Despite the complete volume reduction, adopting a judicious choice of twisted boundary condition the color degrees of freedom provide an effective torus size via the relation $\ell = aL= a\sqrt{N}$ by virtue of large-$N$ volume independence. If the twisted gradient flow coupling is defined in this framework, its running is thus achieved varying $N$. This strategy has been already employed to determine the large-$N$ step-scaling function in the TGF scheme in~\cite{GarciaPerez:2014azn}. Thus, it could in principle also be employed to efficiently determine $\LambdaMS(N\to\infty)$.\\

\FloatBarrier

\acknowledgments
It is a pleasure to thank Massimo D'Elia and Alberto Ramos for collaboration in the initial stages of this project. We are also grateful to Antonio Gonz\'alez-Arroyo for many fruitful discussions. This work is partially supported by the Spanish Research Agency (Agencia Estatal de Investigaci\'on) through the grant IFT Centro de Excelencia Severo Ochoa CEX2020-001007-S and, partially, by the grants PID2021-127526NB-I00 and PID2024-160152NB-I00, all funded by MCIN/AEI/10.13039/501100011033. This work has also been partially supported by the U.S. Department of Energy, Office of Science, Office of Nuclear Physics under Contract No.~DE-SC0012704 and by the Scientific Discovery through Advanced Computing (SciDAC) award ``Fundamental Nuclear Physics at the Exascale and Beyond'' and the Topical Collaboration in Nuclear Theory ``Heavy-Flavor Theory (HEFTY) for QCD Matter''. Numerical calculations have been performed on the Leonardo machine at CINECA, based on the agreement between INFN and CINECA, under the projects \texttt{INF24\_npqcd}, \texttt{INF25\_npqcd} and \texttt{INF26\_npqcd}, on the Finisterrae~III cluster at CESGA (Centro de Supercomputaci\'on de Galicia), and on the Drago HPC cluster, operated by the Scientific Computing Area of SGAI-CSIC.

\section*{Data availability}
All data supporting the findings of this article are openly available in Appendix~\ref{appendix:rawdata} and Appendix~\ref{appendix:fits}. Further data available upon reasonable request.

\renewcommand*{\bibfont}{\footnotesize}
\bibliographystyle{apsrev4-2}
\bibliography{biblio}

\clearpage
\onecolumngrid
\appendix
\section*{Appendix}
\section{Raw data}\label{appendix:rawdata}

\noindent We present the full collection of raw data entering the calculation of the step-scaling function for $N=5,8$ in \cref{tab:rawdata}. Raw data for \SU{3} can be found in the previous papers~\cite{Bribian:2021cmg,Bonanno:2024nba}.

\begin{table}[!htb]
	\tiny
	\begin{subtable}{0.48\textwidth}
		\begin{tabular}{cccccc}
			\midrule
			$b$ & $L=10$ & $L=20$ & $L=30$ & $L=40$ & $L=60$\\
			\midrule
			0.3300    &       $291.36(28)^{*}$      &               &               &               &               \\
			0.3350   &       $129.36(67)^{*}$      &               &               &               &               \\
			0.3400    &       $49.82(18)^{*}$       &       $167.97(43)^{*}$      &               &               &               \\
			0.3450   &    $32.71(11)^{*}$       &       $110.00(30)^{*}$      &       $268.23(65)^{*}$      &               &               \\
			0.3499  &     $24.809(87)^{*}$      &       $78.21(22)^{*}$       &               &               &               \\
			0.3550   &    $20.036(60)^{*}$      &       $57.45(17)^{*}$       &   $134.47(39)^{*}$          &     $256.17(75)^{*}$         &               \\
			0.3597   &      $17.186(14)^{*}$      &     $44.863(34)^{*}$         &            &            &               \\
			0.3604  &     $16.862(45)^{*}$      &  $43.03(14)^{*}$       &               &               &               \\
			0.3630   &    $15.678(40)^{*}$      &  $38.25(13)^{*}$       &       $82.79(41)^{*}$       &       $156.88(61)^{*}$      &       $380.50(210)^{*}$     \\
			0.3658  &               &   $33.91(11)^{*}$       &               &       $133.57(44)^{*}$      &               \\
			0.3680   &    $13.975(33)^{*}$      &     $30.78(11)^{*}$       &       $63.06(18)^{*}$       &       $116.65(30)^{*}$      &       $286.99(149)^{*}$     \\
			0.3705  &   $13.333(29)^{*}$      &   $27.93(11)^{*}$       &               &               &               \\
			0.3720   &  $12.971(27)^{*}$      &   $25.99(10)^{*}$       &       $51.55(15)^{*}$       &       $92.77(25)^{*}$       &       $229.07(149)^{*}$     \\
			0.3766  &               &   $22.081(81)^{*}$      &               &       $73.00(25)^{*}$       &               \\
			0.3778  &               &               &   $39.71(14)^{*}$       &               &       $166.52(84)^{*}$      \\
			0.3790   &    $11.542(22)^{*}$      &       $20.352(77)^{*}$      &   $37.64(14)^{*}$       &       $64.67(26)^{*}$       &       $155.67(73)^{*}$      \\
			0.3821  &       10.989(22)      &       18.648(74)      &               &               &               \\
			0.3868  &               &    $16.354(11)^{*}$      &               &       $44.855(92)^{*}$      &               \\
			0.3876  &               &    $16.062(45)^{*}$      &               &       $43.34(22)^{*}$      &               \\
			0.3889  &               &               &    $25.58(11)^{*}$       &               &       $91.37(80)^{*}$       \\
			0.3930   &       9.536(16)       &       14.300(47)      &               &               &               \\
			0.3986  &               &    $12.927(32)^{*}$      &               &    $27.63(17)^{*}$       &               \\
			0.4000     &               &               &    $17.984(76)^{*}$      &               &       $53.65(59)^{*}$       \\
			0.4040   &       8.475(14)       &       11.777(28)      &               &               &               \\
			0.4041   &             &             &    $16.237(14)^{*}$         &               &   $45.41(45)^{*}$         \\
			0.4097  &               &       10.869(22)      &               &       19.31(11)       &               \\
			0.4111  &               &               &  $14.004(44)^{*}$      &               &   $34.14(53)^{*}$       \\
			0.4149  &       7.640(12)       &       10.136(22)      &               &               &               \\
			0.4208  &               &       9.504(17)       &               &       14.769(64)      &               \\
			0.4222  &               &               &    $11.613(28)^{*}$      &               &   $22.38(28)^{*}$       \\
			0.4255  &               &       9.022(15)       &               &       13.547(48)      &               \\
			0.4300    &               &       8.563(14)       &               &       12.534(41)      &               \\
			0.4319  &               &       8.426(14)       &               &       12.272(42)      &               \\
			0.4333  &               &               &       10.044(19)      &               &       16.82(21)       \\
			0.4337  &               &               &       9.972(19)       &               &       16.53(27)       \\
			0.4381  &               &               &       9.484(17)       &               &       14.86(18)       \\
			0.4404  &               &       7.817(12)       &               &       10.720(27)      &               \\
			0.4426  &       6.1503(88)      &       7.614(12)       &               &               &               \\
			0.4430   &               &               &       9.002(16)       &               &       13.69(16)       \\
			0.4444  &               &               &       8.866(15)       &               &       13.168(94)      \\
			0.4484  &               &               &       8.520(15)       &               &       12.441(96)      \\
			0.4500    &       5.8636(83)      &       7.173(11)       &               &       9.566(21)       &               \\
			0.4549  &               &               &       8.018(14)       &               &       11.344(78)      \\
			0.4596  &               &       6.666(10)       &               &       8.670(18)       &               \\
			0.4600    &       5.5008(76)      &       6.652(10)       &               &       8.629(18)       &               \\
			0.4622  &               &               &       7.516(12)       &               &       10.232(99)      \\
			0.4702  &       5.1722(71)      &       6.1650(100)     &               &               &               \\
			0.4722  &               &               &       6.932(11)       &               &       9.181(57)       \\
			0.4800    &       4.9237(67)      &       5.8190(94)      &               &       7.206(14)       &               \\
			0.4873  &               &       5.5344(80)      &               &       6.795(11)       &               \\
			0.4979  &       4.4910(60)      &       5.2020(73)      &               &               &               \\
			0.5000     &       4.4503(59)      &       5.1418(76)      &       5.7023(82)      &               &       7.129(38)       \\
			0.5150   &               &       4.7411(64)      &               &       5.6532(98)      &               \\
			0.5256  &       3.9728(51)      &       4.5030(69)      &               &               &               \\
			0.5278  &               &               &       4.8763(67)      &               &       5.913(30)       \\
			0.5400    &       3.7474(47)      &       4.2159(57)      &               &               &               \\
			0.5427  &               &       4.1637(61)      &               &       4.8388(80)      &               \\
			0.5534  &       3.5592(45)      &       3.9758(60)      &               &               &               \\
			0.5556  &               &               &       4.2557(59)      &               &       4.972(22)       \\
			0.5705  &               &       3.7126(49)      &               &       4.2417(98)      &               \\
			0.5812  &       3.2284(41)      &       3.5650(71)      &               &               &               \\
			0.5833  &               &               &       3.7887(50)      &               &       4.356(19)       \\
			0.5982  &               &       3.3608(43)      &               &       3.7630(59)      &               \\
			0.6111  &               &               &       3.4043(44)      &               &       3.853(16)       \\
			0.6367  &       2.7237(33)      &       2.9542(38)      &               &               &               \\
			0.6538  &               &       2.8108(35)      &               &       3.0876(45)      &               \\
			0.6600    &       2.5513(32)      &       2.7636(35)      &               &               &               \\
			0.6667  &               &               &       2.8575(36)      &               &       3.176(14)       \\
			0.6923  &       2.3559(28)      &       2.5354(41)      &               &               &               \\
			0.7093  &               &       2.4202(34)      &               &       2.6289(39)      &               \\
			0.7222  &               &               &       2.4558(31)      &               &       2.679(12)       \\
			0.7478  &       2.0811(25)      &       2.2122(30)      &               &               &               \\
			0.7649  &               &       2.1277(27)      &               &       2.2893(32)      &               \\
			0.7778  &               &               &       2.1546(26)      &               &       2.3206(78)      \\
			0.8034  &       1.8596(22)      &       1.9666(27)      &               &               &               \\
			0.8202  &               &       1.8965(23)      &               &       2.0187(26)      &               \\
			0.8333  &               &               &       1.9242(23)      &               &       2.0418(76)      \\
			\bottomrule
		\end{tabular}
		\caption{Collection of \SU{5} data.}
	\end{subtable}%
	\tiny
	\centering
	\begin{subtable}{0.48\textwidth}
		\begin{tabular}{cccccc}
			\midrule
			$b$ & $L=16$ & $L=32$ & $L=48$ & $L=64$ & $L=96$\\
			\midrule
			0.3516  &       $69.94(15)^{*}$       &               &               &               &               \\
			0.3555  &       $54.09(14)^{*}$       &               &               &               &               \\
			0.3555  &       $53.72(13)^{*}$       &               &               &               &               \\
			0.3594  &     $43.08(11)^{*}$       &               &               &               &               \\
			0.3633  &     $35.25(10)^{*}$       &               &               &               &               \\
			0.3657  &    $31.59(10)^{*}$       &       $115.28(21)^{*}$      &               &               &               \\
			0.3666  &               &               &               &               &               \\
			0.3766  &   $20.355(68)^{*}$      &       $62.93(17)^{*}$       &               &               &               \\
			0.3777  &               &       $59.30(16)^{*}$       &               &               &               \\
			0.3778  &               &               &       $137.46(48)^{*}$      &               &               \\
			0.3876  &    $15.060(43)^{*}$      &    $37.38(13)^{*}$       &               &               &               \\
			0.3888  &               &   $35.51(14)^{*}$       &               &       $140.08(88)^{*}$      &               \\
			0.3891  &    $14.530(39)^{*}$      &  $35.46(13)^{*}$       &       $75.26(47)^{*}$       &               &               \\
			0.3986  &  $12.168(25)^{*}$      &   $24.40(12)^{*}$       &               &               &               \\
			0.4000     &               &   $23.23(11)^{*}$       &       $45.11(33)^{*}$       &       $77.50(85)^{*}$       &       $189.49(229)^{*}$     \\
			0.4096  &       10.396(19)      &       17.317(67)      &               &               &               \\
			0.4109  &               &               &   $28.85(26)^{*}$       &               &       $104.59(203)^{*}$   \\
			0.4111  &               &    $16.745(61)^{*}$      &               &       $46.90(59)^{*}$       &               \\
			0.4207  &       9.086(16)       &       13.631(39)      &               &               &               \\
			0.4219  &               &               &   $20.02(20)^{*}$       &               &       $66.06(231)^{*}$      \\
			0.4222  &               &   $13.305(34)^{*}$      &               &    $31.45(49)^{*}$       &               \\
			0.4318  &       8.115(14)       &       11.375(25)      &               &               &               \\
			0.4333  &               &    $11.151(25)^{*}$      &               &    $20.98(52)^{*}$       &               \\
			0.4336  &               &               &   $14.851(95)^{*}$      &               &   $40.45(426)^{*}$      \\
			0.4337  &               &    $11.089(23)^{*}$      &               &      $20.25(59)^{*}$       &               \\
			0.4308   &               &       10.455(21)      &               &       18.26(39)       &               \\
			0.4430   &               &       9.859(18)       &               &       16.09(11)       &               \\
			0.4444  &               &       9.668(18)       &               &       15.671(85)      &               \\
			0.4445  &       7.253(11)       &       9.662(18)       &   $12.317(80)^{*}$      &               &       $26.03(181)^{*}$      \\
			0.4484  &               &       9.260(16)       &               &       14.306(60)      &               \\
			0.4548  &               &       8.651(15)       &               &       12.753(46)      &               \\
			0.4595  &       6.4649(98)      &       8.268(14)       &               &               &               \\
			0.4622  &               &       8.028(13)       &               &       11.352(70)      &               \\
			0.4722  &               &       7.393(12)       &       8.750(33)       &       10.000(22)      &       13.19(27)       \\
			0.4873  &       5.3876(77)      &       6.5671(97)      &               &               &               \\
			0.5000     &               &       6.0077(90)      &       6.822(20)       &       7.629(14)       &       9.17(11)        \\
			0.5150   &       4.6315(62)      &       5.4539(78)      &               &               &               \\
			0.5277  &               &       5.0830(75)      &               &       6.170(11)       &               \\
			0.5281  &               &               &       5.635(17)       &               &       7.000(63)       \\
			0.5427  &       4.0796(54)      &       4.6935(63)      &               &               &               \\
			0.5555  &       3.8811(51)      &       4.4163(61)      &       4.834(13)       &               &       5.773(48)       \\
			0.5555  &               &       4.4126(62)      &               &       5.2022(68)      &               \\
			0.5705  &       3.6395(46)      &       4.1257(55)      &               &               &               \\
			0.5833  &               &       3.9113(53)      &               &       4.5009(53)      &               \\
			0.5836  &               &               &       4.237(11)       &               &       4.889(38)       \\
			0.5982  &       3.2893(42)      &       3.6812(47)      &               &               &               \\
			0.6109  &               &               &       3.7499(94)      &               &       4.343(25)       \\
			0.6111  &               &       3.5077(47)      &               &       3.9664(47)      &               \\
			0.6538  &       2.7648(35)      &       3.0356(40)      &               &               &               \\
			0.6664  &               &               &       3.1016(75)      &               &       3.445(16)       \\
			0.6666  &               &       2.9177(38)      &               &       3.2252(36)      &               \\
			0.7093  &       2.3896(28)      &       2.5824(32)      &               &               &               \\
			0.7219  &               &               &       2.6184(66)      &               &       2.895(13)       \\
			0.7222  &               &       2.4986(31)      &               &       2.7230(30)      &               \\
			0.7648  &       2.1020(25)      &       2.2513(28)      &               &               &               \\
			0.7777  &               &       2.1875(28)      &               &       2.3559(31)      &               \\
			0.7781  &       2.0486(23)      &       2.1858(27)      &       2.2817(54)      &               &       2.457(16)       \\
			0.8202  &       1.8817(22)      &       1.9967(23)      &               &               &               \\
			0.8333  &               &       1.9434(23)      &               &       2.0665(72)      &               \\
			0.8336  &               &               &       2.0141(47)      &               &       2.1773(83)      \\
			0.8891  &               &               &       1.8091(41)      &               &       1.9209(83)      \\
			0.9445  &               &               &       1.6525(38)      &               &       1.7342(76)      \\
			\bottomrule
		\end{tabular}
		\caption{Collection of \SU{8} data.}
	\end{subtable}
	\caption{Collection of raw data for the TGF ’t Hooft coupling as a function of $b$ and $L$. Data marked by an asterisk have been used for the initial step $\muhad \to 2\muhad$, while the others for the computation of the continuum step-scaling function from $2\muhad$ onward.}
	\label{tab:rawdata}
\end{table}

\clearpage
\section{Parameters of the continuum step-scaling function}\label{appendix:fits}

For completeness, we report in \cref{tab:fit_parameters} the best-fit parameters $\hat{p}_n$ of the continuum parameterization of the inverse step-scaling function in \cref{eq:step_global_fit_var} with $d_1=4$:
\be
\nonumber
\frac{1}{u(\sigma)} - \frac{1}{\sigma} = 2b_0\log 2 +2b_1\sigma\log 2 +\hat p_2 \sigma^2 +\hat p_3 \sigma^3 +\hat p_4 \sigma^4 \, .
\ee
The corresponding covariance matrices are given in \cref{tab:cov_SUN}. 

\begin{table}[h!]
	\centering
	\begin{tabular}{cccc}
		\toprule
		& \SU{3} & \SU{5} & \SU{8} \\
		\midrule
		$\hat p_2$
		& $-3.58241654\times10^{-5}$
		& $-6.03717851\times10^{-5}$
		& $-3.70751543\times10^{-5}$ \\
		
		$\hat p_3$
		& $ 2.28461884\times10^{-6}$
		& $ 3.81535245\times10^{-6}$
		& $ 5.09045437\times10^{-7}$ \\
		
		$\hat p_4$
		& $-3.58559744\times10^{-8}$
		& $-3.16998328\times10^{-8}$
		& $ 8.06461112\times10^{-8}$ \\
		\bottomrule
	\end{tabular}
	\caption{Best-fit parameters of the continuum step-scaling function for each value of $N$.}
	\label{tab:fit_parameters}
\end{table}

\begin{table}[h!]
	\centering
	\begin{tabular}{ccccc}
		\toprule
		$N$ & & $\hat p_2$ & $\hat p_3$ & $\hat p_4$ \\
		\midrule
		\multirow{3}{*}{3}
		& $\hat p_2$
		& $ 2.78928259\times10^{-10}$
		& $-4.34235470\times10^{-11}$
		& $ 1.61945139\times10^{-12}$ \\
		
		& $\hat p_3$
		& $-4.34235470\times10^{-11}$
		& $ 6.91409571\times10^{-12}$
		& $-2.62047961\times10^{-13}$ \\
		
		& $\hat p_4$
		& $ 1.61945139\times10^{-12}$
		& $-2.62047961\times10^{-13}$
		& $ 1.00550851\times10^{-14}$ \\
		\midrule
		\addlinespace[0.5em]
		\multirow{3}{*}{5}
		& $\hat p_2$
		& $ 4.32759232\times10^{-10}$
		& $-7.25014431\times10^{-11}$
		& $ 2.94447591\times10^{-12}$ \\
		
		& $\hat p_3$
		& $-7.25014431\times10^{-11}$
		& $ 1.24481228\times10^{-11}$
		& $-5.15239698\times10^{-13}$ \\
		
		& $\hat p_4$
		& $ 2.94447591\times10^{-12}$
		& $-5.15239698\times10^{-13}$
		& $ 2.16593102\times10^{-14}$ \\
		\midrule
		\addlinespace[0.5em]
		\multirow{3}{*}{8}
		& $\hat p_2$
		& $ 6.06883333\times10^{-10}$
		& $-9.78552849\times10^{-11}$
		& $ 3.79591332\times10^{-12}$ \\
		
		& $\hat p_3$
		& $-9.78552849\times10^{-11}$
		& $ 1.62858876\times10^{-11}$
		& $-6.46082723\times10^{-13}$ \\
		
		& $\hat p_4$
		& $ 3.79591332\times10^{-12}$
		& $-6.46082723\times10^{-13}$
		& $ 2.60543219\times10^{-14}$ \\
		\bottomrule
	\end{tabular}
	\caption{Covariance matrices of the best-fit parameters of the continuum step-scaling function for each value of $N$.}
	\label{tab:cov_SUN}
\end{table}

\end{document}